\documentclass[usenatbib]{mn2e}
\usepackage{graphicx}
\usepackage{amssymb}
\newcommand{\fracb}[2]{\left(\frac{#1}{#2}\right)}

\newcommand{\mean}[1]{\langle{#1}\rangle}

\newcommand{\text}[1]{\quad\mbox{#1}\quad}

\newcommand{\aap}{A\&A}

\title[Shock dissipation]
{Shock Dissipation in Magnetically Dominated Impulsive Flows}

\author[Komissarov S.S.]
{
Serguei S.~Komissarov 
\\
Department of Applied Mathematics, The University of Leeds, Leeds, LS2 9GT, UK\\
E-mail: serguei@maths.leeds.ac.uk
}

\begin{document}
\date{Received/Accepted}
\maketitle

\begin{abstract} 
The idea that cosmic relativistic jets are magnetically driven Poynting-dominated flows 
has many attractive features but also some problems. 
One of them is the low efficiency of shock dissipation in highly 
magnetized plasma. Indeed, the observations of gamma ray bursts (GRBs) and their 
afterglow emission indicate very high radiative efficiency of relativistic jets 
associated with these phenomena. We have revisited the issue of shock dissipation 
and emission and its implications for the internal shock model of the prompt GRB 
emission and studied it in the context of impulsive Poynting-dominated flows.  
Our results show that unless the magnetization of GRB jets is extremely high, 
$\sigma > 100$ in the prompt emission zone, the magnetic model may still be compatible 
with the observations.  First, for $\sigma\simeq 1$ the dissipation efficiency of 
fast magnetosonic shock is still quite high, $\sim 30\%$. Second, the main effect 
of reduced dissipation efficiency is merely an increase in the size of the dissipation 
zone and even for highly magnetised GRB jets this size may remain below the external 
shock radius, provided the central engine can emit magnetic shells on the time scale 
well below the typical observed variability scale of one second.  
Our analytical and numerical 
results suggest that magnetic shells begin strongly interact with each other 
well before they reach the coasting radius. As the result, the impulsive jet in 
the dissipation zone is best described not as a collection of shells but as a continuous
highly magnetised flow with a high amplitude magnetosonic wave component.           
How exactly the dissipated wave energy is distributed between the radiation and 
the bulk kinetic energy of radial jets depends on the relative rates of radiative and 
adiabatic cooling.  In the fast radiative cooling regime, the corresponding radiative 
efficiency can be as high as the wave contribution to their energy budget, independently 
of the magnetization. Moreover, after leaving the zone of prompt emission the
jet may still remain Poynting-dominated, leading to weaker emission from the reverse
shock compared to non-magnetic models.
Energetically sub-dominant weakly magnetized ``clouds'' in otherwise strongly magnetised 
jets may significantly increase the overall efficiency of the shock dissipation. 

\end{abstract}


\section{Introduction}
\label{sec:introduction}

Various astronomical observation indicate, directly or indirectly, the existence of 
highly relativistic outflows in a variety of cosmic phenomena, such as active galaxies, 
pulsar wind nebulae, X-ray binary stars, and Gamma Ray Bursts (GRBs).  
Although the origin of these flows is still a subject 
of debate, especially in the case of GRBs, and requires further investigation, 
so far we have identified only one mechanism of jet production which may operate 
in all these very diverse environments -- the magnetic mechanism. This is one of the
reasons why this mechanism has attracted so much attention in recent years.   

It has been shown that magnetic fields can not only tap the rotational energy of a
massive rotator, placed in the ``hart'' of ``cosmic jet engines'' in this model, but 
also accelerate and collimate outflows. In fact, it has been shown that   
in the case of relativistic steady-state jets, their magnetic acceleration and 
collimation go hand-in-hand and for this reason this version of magnetic mechanism 
is called {\it the collimation acceleration mechanism}. In order to ensure efficient magnetic 
acceleration, the jet opening angle should be small compared to the jet Mach 
angle associated with fast magnetosonic waves \citep{KVKB09}. In the small 
angle approximation, the Mach angle
$$
   \theta_{\rm M} \simeq \frac{1}{M}\, ,
$$
where $M$ is the fast magnetosonic Mach number. 
In the case of predominantly asymuthal magnetic field this 
number can be estimated using Eq.\ref{mach} and the acceleration 
condition reads
$$
    \sigma \gtrsim (\gamma_{\rm j} \theta_{\rm j})^2, 
$$
where $\sigma=B^2/\rho c^2$. In order to understand where this limitation comes from, 
consider a freely expanding conical jet. 
If its magnetic field is predominantly azimuthal than
the magnetic freezing yields $B\propto r^{-1}$ and the magnetic energy
${\cal E}_{\rm m}\propto r^2 B^2 \propto r^0$ is not utilised to accelerate the
flow. The flow geometry has to deviate from the conical one in order for the
magnetic acceleration to operate and this requires efficient causal communication
across the flow \citep[e.g.][]{K11}. In fact, the above causality condition is a 
bit too strict and the acceleration may proceed, though at a much lower logarithmic 
rate, even after it is no longer satisfied. However, the above constraint on $\sigma$ 
remains valid up to a factor of few \citep{L09}.        
For AGN jets with their inferred $\gamma_{\rm j}\sim 10$ and observed $\theta_{\rm j}\simeq 0.1$ 
this yields $\sigma \gtrsim 1$, whereas for the commonly accepted parameters of GRB jets,
$\gamma_{\rm j} \ge 100$ and $\theta_{\rm j}\simeq 0.1$, the last equations implies $\sigma\gg 1$.

The observed variability of the emission produced in these jets has been 
traditionally associated with strong shock waves, driven into the jets by 
their unsteady central engines \citep[see the review by][]{Piran}. 
Their observed bright knots and 
spots have also been attributed to such internal shocks.  
Indeed, shocks are generally known as places of effective dissipation of 
kinetic energy and acceleration of relativistic electrons, 
responsible for the non-thermal emission observed in many astrophysical objects. 
However, in the case of relativistic magnetized flows this interpretation 
encounters significant problems, particularly in the case of GRB jets 
\citep[e.g.][]{NKT11}. 

Indeed, the dissipation efficiency of relativistic shocks in highly magnetized plasma 
is rather low. This was widely accepted already after the pioneering work 
by \citet{KC84}. To be more accurate, this statement is concerned with fast 
magnetosonic shocks only. Slow magnetosonic shocks can still have very high 
dissipation efficiency, because at such shocks the magnetic energy can be 
dissipated as well \citep{Lyub05}. However, 
fast shocks are much more readily produced, usually via collisions, 
whereas formation of slow shock requires some rather special conditions.     


In contrast, the observations indicate that the jet radiative efficiency, 
$\eta_{\rm r}$, defined as the fraction of jet energy eventually converted into 
radiation (usually non-thermal), can be quite high. For example, the observations of 
the GRB afterglows imply that the radiative efficiency of their jets is often in 
excess of $10\%$ and sometimes may even reach 
$90\%$ \citep{PK02,Y03,GKP06,Z07a}.  According to the more recent study of {\it Swift}
GRBs the situation is even more dramatic, with the mean radiative efficiency 
around $90\%$ \citep{W07}.  This difficulty of the shock model has forced  
many theorists to start looking for alternative models involving direct dissipation of 
magnetic energy associated with the magnetic reconnection 
\citep[e.g.][]{DS02,ZY11,MU10,Lyub10}. 
However, these models still remain at a rather rudimentary level of development due 
to difficulties of their own.   

{\bf
In most of the previous theoretical studies of magnetized relativistic 
jets it was assumed that they were more or less homogeneous, just for sake 
of simplicity. The strong observed 
variability and the complex observed structure of some relativistic jets, where high 
resolution images are available, suggest that this may be not a very realistic 
assumption. Following the early work by \citet{C95}, a number of recent papers 
explored the implications of highly intermittent jet production on its 
dynamics \citep{GKS10,L10,LL10,L11,G11a,G11b}. They have concluded that longitudinal
expansion of highly magnetised plasma shells may result in efficient conversion 
of the Poynting flux into the bulk kinetic energy of the shells and strong 
reduction of their magnetisation. However, in those papers only the dynamics of 
a single shell was studied in details, whereas the case of an impulsive jet 
composed of many such shells was subjected to a much more speculative analysis.      
The main goal of our study was to reduce this imbalance.      

As a first step in studying of the multiple shell case one may assume that 
the gaps between them are empty. However, if the jet engine does indeed operate 
in an impulsive fashion then external plasma, presumably of much lower magnetization, 
is likely to fill the gaps during quiescent periods. Then each time the jet is 
reborn it has to push this plasma aside. Provided the jet is sufficiently 
powerful, this can be done quite efficiently by the bow-shock developing at the 
jet head \citep[e.g.][]{KF98}. Farther out, where the 
distance between shells becomes comparable to the jet radius, the relativistic 
effects make impossible for the external plasma to enter the gaps \citep{LL10}.     
At the same time it becomes impossible for the entrained plasma to leave the 
gaps as it is forced to remain within the cone of the half-opening angle 
$1/\gamma_{\rm j}$. Thus, it seems quite plausible that some of the external 
plasma will remain in the gaps and become part of the jet, though it is still 
rather difficult to quantify the effectiveness of this mass-loading process at 
present.  

On one hand, loading Poynting-dominated jets with weakly magnetised clouds 
complicates the problem. On the other hand, this may play a very important role 
in their physics. As far as the radiative efficiency is concerned, these clouds 
could be the locations there most of the shock dissipation and emission 
takes place. Very much in the same way as in the model of the afterglow emission, 
where a magnetic piston drives the so-called external shock wave 
through weakly magnetised interstellar medium \citep[e.g.][]{LB03,ZK05,L06}.         
Obviously, these clouds may have to be ``excited'' many times before a significant 
fraction of magnetic energy is radiated. 
}

A similar repetitive ``pumping'' action has been investigated by \citet{KS01} 
in the case of unmagnetized highly variable jets. As individual shells 
(portions of the jet moving with very different Lorentz factors) collide and heat-up, 
only a fraction of the dissipated energy is radiated. The rest of it remains initially
in the form of heat, but later, when the shells begin to expand, this heat 
is converted back into the kinetic energy of relative motion. When another
collision occurs, a fraction of this energy is dissipated and radiated
again and so on. The process continues until the shocks become very
weak.  \citet{KS01} have demonstrated that this way the radiative
efficiency can be increased up to $60\%$, even if during each
individual collision only $10\%$ of the dissipated energy is
radiated. However in order to achieve this, they required very strong 
variations of the jet Lorentz factor, with uniform distribution of 
$\log_{10}\gamma$ between 1 and 4.
Since in our case the energy behind this dynamics is of magnetic nature, 
the name ``magnetic pump'' springs to mind. 
In fact, each time the shock-heated gap plasma expands and its components, 
which are unable to cool radiatively, cool adiabatically their  
thermal energy is returned back to the ``pump'' and recycled.  

The whole problem of impulsive jet dynamics from its production to its 
interaction with the interstellar matter is still prohibitively complex. 
In order to make progress, we will consider much simpler problems hoping to 
elucidate some of its important aspects.  We start with the issue of the dissipation 
efficiency of fast magnetosonic shocks in highly magnetized plasma as we feel need 
to clarify few important points. This is done in Section \ref{sec:shock-eff} and in 
Appendix~\ref{appendix-1}. Then we consider strictly periodic one-dimensional flows 
in slab geometry in the framework of one fluid MHD with simple polytropic equation of state. 
In Section~\ref{sec:chamber} we study oscillations developing in a periodic train 
of initially stationary magnetic shells, as they expand, collide and radiatively cool. 
Then we consider moving trains with initially empty gaps, following  
\citet{GKS10}, first in the adiabatic regime (Section~\ref{sec:adiabatic}) and then 
in the regime of fast radiative cooling (Section~\ref{sec:FCF}). 
Finally, we study the dynamics of a moving train with gaps filled with weakly 
magnetised plasma from the start in the fast radiative cooling regime 
(Section~\ref{sec:FCF-FG}). The results and their astrophysical implications are 
discussed in Section~\ref{sec:discussion}. 
Our conclusions are listed in Section~\ref{sec:conclusions}.

Throughout most of the paper we use the Heaviside units, where $c=1$ and $B/\sqrt{4\pi}\to B$, 
but in the Discussion we reintroduce the speed of light.

\section{Dissipation efficiency of perpendicular fast magnetosonic shocks.}
\label{sec:shock-eff}

Here we consider only the perpendicular relativistic shocks, where the flow velocity 
is perpendicular to the shock front. In addition we assume that the magnetic field is 
parallel to the shock front. Thus, we restrict our attention to purely  one-dimensional flow 
directed perpendicular to the magnetic field. This constraint prohibits slows 
magnetosonic waves of any kind and the only non-trivial shocks solutions are 
the fast magnetosonic ones. 
Such shocks were first studied by \citet{KC84} in application to the termination 
shocks of pulsar winds. In order to simplify the shock equations, they only 
considered the case of cold upstream flow, ultrarelativistic shock speed, and 
high shock strength, in the sense that the downstream Lorentz factor is much lower 
compared to the upstream one. All these additional constrains are justified 
in the case of pulsar wind nebulae. The general case of magnetosonic shocks was 
analysed by \citet{MA87} and \citet{AC88}.  Later, \citet{ZK05} expanded the analysis 
of \citet{KC84} by allowing variable shock strength, which they described by 
the Lorentz factor of the relative motion between the upstream and downstream 
states, $\gamma_{12}$. Although this parameter can indeed be used to describe 
the shock strength, the more traditional parameter, unanimously accepted in 
the non-relativistic hydrodynamics and MHD, is the shock Mach number. 
The proper relativistic definition of Mach number with respect to the wave mode 
of speed $c_{\rm m}$ in the fluid frame is

\begin{equation}
   M = \frac{\gamma v}{\gamma_{\rm m} c_{\rm m}}, 
\label{mn}
\end{equation} 
where $\gamma_{\rm m}$ is the Lorentz factor corresponding to the wave speed $c_{\rm m}$ 
and $\gamma$ is the Lorentz factor corresponding to the flow speed $v$ \citep{K80}.            
\begin{figure*}
\includegraphics[width=80mm,angle=-90]{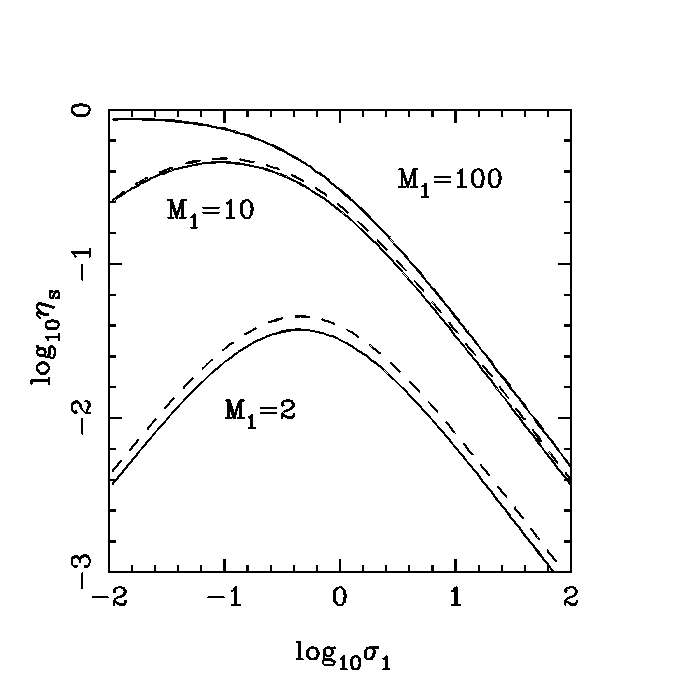}
\includegraphics[width=80mm,angle=-90]{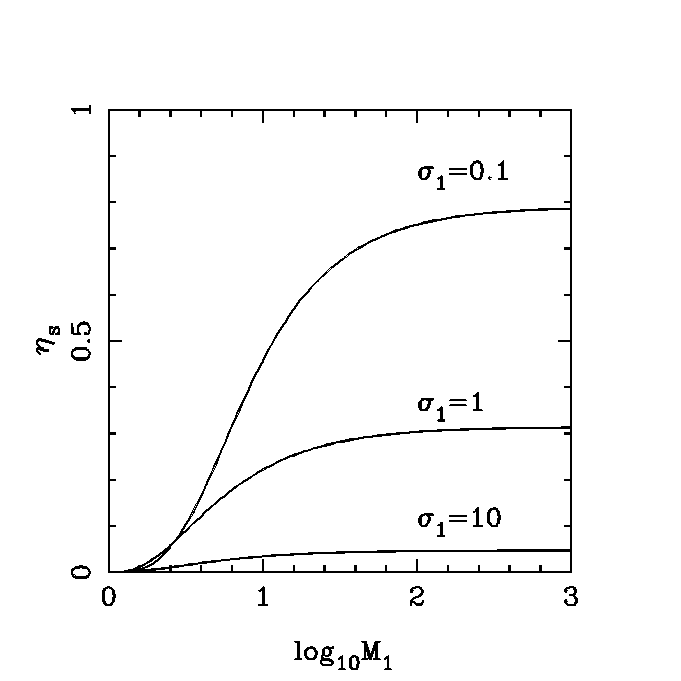}
\caption{ {\it Left panel:} Dissipation efficiency of perpendicular
shock as a function of the magnetisation parameter $\sigma$ of the 
upstream state for three different values of the shock 
fast magnetosonic Mach number. The solid lines show the solutions 
for the polytropic equation of state with $\Gamma=4/3$ ($\kappa=4$). 
The dashed lines show the solutions for the electron-positron 
Synge gas.
{\it Right panel:} Dissipation efficiency of perpendicular
shock as a function of the shock fast magnetosonic Mach number $M_1$ for 
three different values of the magnetisation parameter $\sigma$ of the
upstream state. The equation of state describes polytropic gas 
with $\Gamma=4/3$ ($\kappa=4$).
}
\label{fig:eff}
\end{figure*}
In the limit of cold plasma, where the thermodynamic pressure, $p$, and  hence 
the sound speed are set to zero, the fast magnetosonic speed is the same in 
all directions

\begin{equation} 
c_{\rm f}^2 = \frac{B^2}{B^2+\rho} = \frac{\sigma}{1+\sigma},
\end{equation}
where $B$ and $\rho$ are the magnetic field strength and the gas rest mass density 
as measured in the fluid frame, and 

\begin{equation}
\sigma = \frac{B^2}{\rho}
\end{equation}
is one of the parameters describing the plasma magnetization. 
For $\gamma\gg 1$ and $\sigma\gg 1$ the fast magnetosonic Mach number is  

\begin{equation}   
   M \simeq \frac{\gamma}{\sqrt{\sigma}}. 
\label{mach}
\end{equation}
In Appendix~\ref{appendix-1} we redo the analysis of perpendicular fast 
shocks of \citet{ZK05} using the Mach number with respect to the fast magnetosonic mode 
as the shock strength parameter. While in general numerical techniques have to be 
used to solve the shock equations, in the limit of high shock Mach number, $M_1\gg 1$, 
and high upstream magnetisation, $\sigma_1\gg 1$, they allow simple approximate 
solution where
\begin{equation}
   \gamma_2 \simeq \sigma_1^{1/2},
\end{equation}
\begin{equation}
   \rho_2 \simeq M_{\rm 1} \rho_1,
\end{equation}
\begin{equation}
   p_2 \simeq \frac{1}{8} \rho_1 M_{\rm 1}^2,
\end{equation}
\begin{equation}
   p_{\rm m,2} \simeq  M_{\rm 1}^2 p_{\rm m,1},
\end{equation}
where $p_{\rm m}$ is the magnetic pressure. Here index ``1'' refers to the 
upstream and index ``2'' to the downstream state.
One can see that for $M_1\gg 1$ there are strong jumps in the rest 
mass density and magnetic pressure as measured in the fluid frame. 
This is what is meant by 
\citet{ZK05}, when they state that high magnetisation does not prevent 
development of strong shocks. On the other hand, if we consider parameters 
measured in the shock frame, which will be indicated by prime, then 
\begin{equation}
   \rho'_2 \simeq \rho'_1 \, ,
\label{rho_p}
\end{equation}
\begin{equation}
   B'_2 \simeq  B'_1 \,
\end{equation} 
where $\rho'$ is the rest mass, not the inertial mass, density. 
Eq.\ref{rho_p} shows that there is no much decrease in the shock frame 
volume occupied by plasma as it crosses the shock -- the large decrease 
in the proper specific volume is almost totally compensated by the reduced
Lorentz contraction. It is this what is meant when shocks in highly 
magnetised medium are often described as weak or weakly compressive.    

One can define the shock dissipation efficiency in many different ways, some more 
meaningful than others. We are interested in the fraction of the total energy 
flux which can be converted into radiation without invoking any additional 
dissipations mechanisms, like the magnetic reconnection, downstream of the shock. 
This suggests to define  the efficiency as
\begin{equation}
\eta_{\rm s}=\frac{F_{\rm t,2}}{F_{\rm tot,2}}
\label{eff_def}
\end{equation}
where $F_{\rm t}=e_{\rm t}\gamma^2 v$ is the thermal energy flux density and 
$F_{\rm tot}=(w+B^2)\gamma^2 v$  is the total energy flux density, $e_{\rm t}$ is 
the thermal energy density and $w=\rho+e_{\rm t}$ is the relativistic enthalpy, 
both defined in the fluid frame. Notice that this definition makes $\eta_{\rm s}$ 
independent on the flow velocity, and hence on the shock speed relative to 
the observer.  For $M_1,\sigma_1\gg1$ we find 
\begin{equation}
\eta_{\rm s}= \frac{1}{2(1+\sigma_1)}
\label{eff}
\end{equation}
(see Appendix~\ref{appendix-1} ). 

\begin{figure*}
\includegraphics[width=60mm,angle=-90]{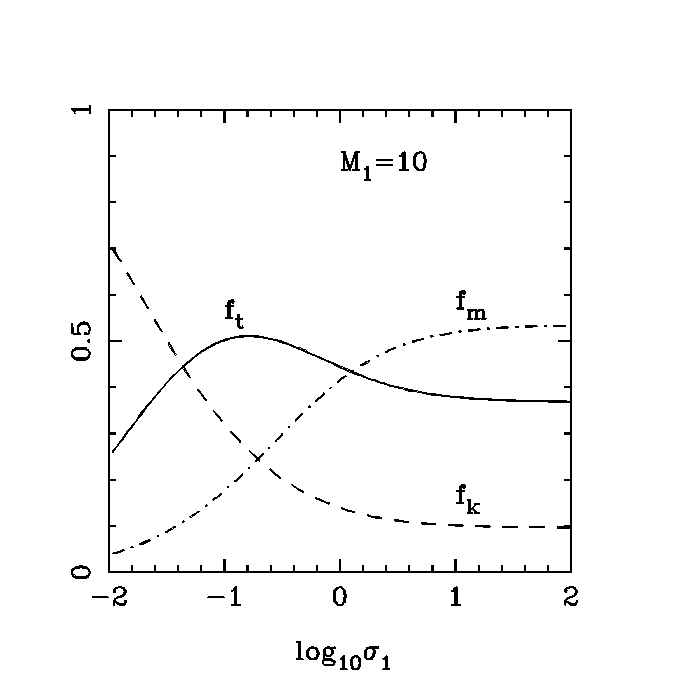}
\includegraphics[width=60mm,angle=-90]{figures/frac-m100.eps}
\includegraphics[width=60mm,angle=-90]{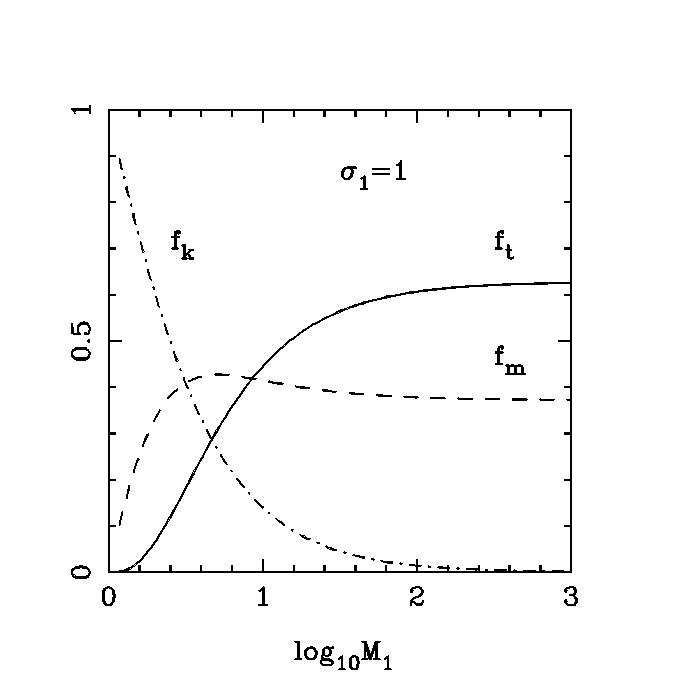}
\caption{ Fractions of the upstream kinetic energy converted at the shock into 
the thermal energy, $f_{\rm t}$, magnetic energy, $f_{\rm m}$, and remaining in the 
kinetic form, $f_{\rm k}$. The left and the middle 
panel show the fractions as functions of the upstream magnetisation for 
$M_1=10$ and $M_1=100$ respectively. The right panel shows them as functions 
of the shock fast magnetosonic Mach number for the upstream magnetisation $\sigma_1=1$.    
}
\label{fig:frac}
\end{figure*}

Equation~(\ref{eff}) shows that for $\sigma_1\gg 1$ only a rather small 
fraction of the flow energy can be dissipated and then radiated away. 
In order to see if the same conclusion applies to shocks in only moderately 
magnetized plasma we solved the shock equations numerically. The results are 
presented in Figures (\ref{fig:eff}) and (\ref{fig:frac}).   
The left panel of Fig.(\ref{fig:eff}) shows the 
dissipation efficiency as a function of $\sigma_1$ for three different values 
of the shock fast magnetosonic Mach number, $M_1=2,10,$ and 100. 
One can see that for $\sigma_1 \gg 1$ the dissipation efficiency 
does indeed decline as $\sigma_1^{-1}$. However, for small magnetization the 
efficiency actually increases with $\sigma_1$. The location of the maximum 
depends on the shock Mach number and for weak shocks is near $\sigma_1=1$. 
However, its magnitude is rather low in this case. As one can see in the right
panel of Fig.\ref{fig:eff} the efficiency monotonically increases with the shock Mach
number. 

The collimation acceleration of magnetic jets may result in the asymptotic 
magnetization $\sigma \simeq 1$. For this reason we presented 
in the right panel of Fig.\ref{fig:eff} the dissipation efficiency as a function 
of the shock Mach number for $\sigma_1$ around this value. For very high Mach
number the efficiency is $\eta_{\rm s} \simeq 0.8,$ 0.3, and 0.05 for 
$\sigma = 0.1,$ 1, and 10 respectively. This shows that for $\sigma_1\simeq 1$ 
the dissipation efficiency can be already reasonably high. Moreover, Eq.\ref{eff} 
gives a rather accurate estimate of the efficiency for $M_1\gtrsim 10$ and 
$\sigma_1 \gtrsim 1$.          

The shock solution depends on the plasma equation of state (EOS) but not 
strongly, at least for the explored range of parameters. In the right panel of 
Fig.\ref{fig:eff} the solid lines show the solution for the polytropic equation 
of state with $\Gamma=4/3$ ($\kappa=4$), whereas the dashed lines show the solution 
for the electron-positron Synge gas, which assumes the same temperature 
relativistic Maxwell distribution for every species \citep{S57}. One can see that 
difference is relatively small, particularly for strong shocks. We also 
experimented with the electron-proton Synge gas and found that 
the solution was even closer to that with the polytropic EOS.

The result (\ref{eff})  has a straightforward 
interpretation. First, only the kinetic energy dissipates at the shock. 
Second, the kinetic energy flux makes only  $1/(1+\sigma_1)$ of the total 
upstream energy flux. Finally, for high $M_1$ approximately one half of the 
kinetic energy dissipates into heat, and approximately one half is 
converted into Poynting flux. This is illustrated in Fig.\ref{fig:frac} which shows    
fractions of the upstream kinetic energy converted at the shock into
the thermal energy, $f_{\rm t}$, magnetic energy, $f_{\rm m}$, and remaining in the
kinetic form, $f_{\rm k}$.

\section{Chamber oscillations}   
\label{sec:chamber} 

Collisions between shells will produce reflected shock waves, similar to 
those created via shock reflection off a conducting wall. So the problem of shell 
interaction appears
analogous to that of a shock bouncing off the ends of a closed tube that contains 
plasma with very diverse magnetization. Each time it crosses the low magnetisation
domain a fraction of its energy is dissipated and radiated away, so the shock 
weakens bit by bit. 

Consider a one-dimensional flow confined within a chamber of 
length $l$. Suppose that initially the chamber is divided into two sections, 
of lengths $l_p$ and $l_g=l-l_p$. The
first section, which we will call the ``pulse'', is filled with uniform
highly magnetized cold plasma and the second section, which will be
referred to as the ``gap'', is uniformly filled with plasma of lower 
magnetization and weaker magnetic field. When comparing this configuration with 
inhomogeneous relativistic jet, one is tempted to identify the length $l$ with the separation 
between two neighbouring shells as measured in the jet frame. 

In the purely electromagnetic version of this problem the gaps are empty and the pulses  
have a uniform distribution of magnetic field with vanishing 
electric field. The solution to this problem involves two identical  
electromagnetic pulses bouncing between the perfectly conducting walls 
of the chamber without decay. When the plasma magnetization is high, 
$\sigma \gg 1$, we expect the MHD solution to be close to the electromagnetic 
one. However, the shock dissipation will gradually damp these 
oscillations. If the radiative cooling of the chamber plasma is indeed very
efficient, it eventually relaxes to an equilibrium with uniform
magnetic field and negligibly small temperature. This allows us to compute 
the total loss of energy from the system, and hence its radiative 
efficiency.  

\subsection{Asymptotic state}

Denote as $\rho_p$, $B_p$, and $M_p$ the
rest mass density, the magnetic field, and the total mass of the pulse respectively, 
and as $\rho_g$, $B_g$, and $M_g$ the corresponding parameters of the gap. It
is convenient to describe the problem by the ratios of lengths, rest
masses, and magnetic energies of the gap and the pulse:
\begin{equation}
  \delta_l = \frac{l_g}{l_p},\quad \delta_m = \frac{M_g}{M_p},\quad
  \delta_e = \frac{B^2_gl_g}{B^2_p l_p}\, .
\label{e1}
\end{equation}

The relativistic magnetization parameter of cold plasma,
$\sigma=B^2/\rho$, gives the ratio of the magnetic and rest mass
energies in the fluid frame. Given this, it makes sense to define the
mean magnetization of plasma in the initial state as
\begin{equation}
  \mean{\sigma}_0 = \frac{1}{c^2}\frac{B_p^2 l_p +B_g^2 l_g}{\rho_p
  l_p +\rho_g l_g} = \frac{\sigma_p+\sigma_g\delta_m}{1+\delta_m} \, .
\label{e3}
\end{equation}
The condition of strong mean magnetization 
constrains the mass fraction of the system. In particular, if the 
gap magnetization is really low, $\sigma_g\ll 1$, this requires 
$ \sigma_p \gg \delta_m$. 

\begin{figure}
\includegraphics[width=80mm]{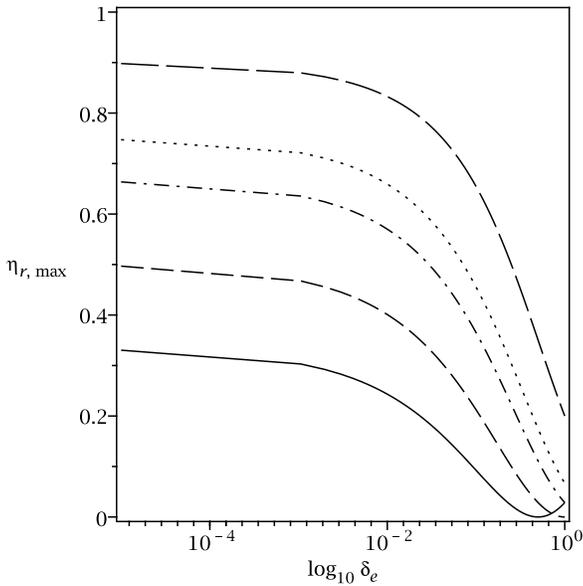}
\caption{Radiative efficiency in the chamber problem for $\delta_l=0.5$ (solid line),
$\delta_l=1$ (short dash line), $\delta_l=2$ (dot line), $\delta_l=3$
(dash-dot line), $\delta_l=9$ (long dash line) as a function of the
energy ratio parameter $\delta_e$.  }
\label{fig:eta_max}
\end{figure}

From the rest mass conservation and the magnetic field freezing  we have 
\begin{equation}
  \rho_g l_g = \tilde{\rho}_g \tilde{l}_g \, , \quad
  \rho_p l_p = \tilde{\rho}_p \tilde{l}_p \, ,
\label{e5}
\end{equation} 
\begin{equation}
  B_g l_g = \tilde{B} \tilde{l}_g \, , \quad
  B_p l_p = \tilde{B} \tilde{l}_p \, ,
\label{e7}
\end{equation} 
where tilde denotes parameters of the equilibrium state, which is reached 
asymptotically for $t\to\infty$. These combine to yeild 
\begin{equation}
  \tilde{B}= B_p \frac{1+\delta_l^{1/2} \delta_e^{1/2}}{1+\delta_l} \, .
\label{e9}
\end{equation}
Using this result one can derive the magnetic energy of the
equilibrium state and hence the radiative efficiency, 
\begin{equation}
  \eta_{\rm r,max} =
  1-\frac{(1+\delta_l^{1/2}\delta_e^{1/2})^2}{(1+\delta_l)(1+\delta_e)} \, ,
\label{e10}
\end{equation}
which is defined here 
as the fraction of the initial magnetic energy converted into radiation. 
This function is shown in Figure~\ref{fig:eta_max}.  One can see that
the radiative efficiency increases with $\delta_l$ and decreases with
$\delta_e$, which has a very simple explanation.  Smaller
$\delta_e$ means stronger expansion of the pulse and hence smaller
magnetic energy remaining in the system after its relaxation.  For
$\delta_e=0$ the efficiency is simply the fraction of 
the volume available for the pulse to fill,
\begin{equation}
\eta_{\rm r,max}=\frac{l_g}{l} \, .
\label{e11}
\end{equation}
The gap plasma however resists the pulse expansion and
its resistance increases with the gap pressure and hence the magnetic energy 
stored in the gap. 

This simple analysis hints that the radiative efficiency of impulsive
magnetically dominated flows can be very high. Moreover, the outcome
does not even depend on the magnetization. Other important properties
of the process however may do. For example the rate of dissipation, and
hence the luminosity. In order to investigate this issue a bit further
we have carried out numerical simulations.

\begin{table}
   \begin{tabular}{l|l|l|l|l|l|l|l|l|l|}
   \hline & $\sigma_p$ & $\sigma_g$ & $\delta_l$ & $\delta_e^{-1}$ &
     $\delta_m$ & $\mean{\sigma}_0\!\!\!$ & $\mean{\sigma}_f\!\!\!$ & $\eta_{\rm
     r,max}\!\!\!$ & $t_{50}$ \\ \hline \hline 
     X & 30 & $10^{-3}\!\!\!$ & 1 & $10^7$ & $10^{-4}\!\!\!\!$ 
                              & 30 & 15 & 0.5 & 3.5 \\ \hline 
     A & 5 & 1.0 & 1 & 5 & 1 & 3 & 1.8 & 0.13 & 3.9 \\ \hline 
     B & 15 & 3.0 & 1 & 5 & 1 & 9 & 5.4 &0.13 & 8.3 \\ \hline 
     C & 15 & 0.1 & 3 & 150 & 1 & 7.6 & 2.1 &
     0.68 & 1.6 \\ \hline D & 30 & 0.1 & 1 & 300 & 1 & 15. & 7.9 &
     0.44 & 4.2 \\ \hline E & 30 & 0.1 & 1 & 30 & 10& 2.8 & 1.6 & 0.32
     & 1.6 \\ \hline F & 30 & 1.0 & 1 & 30 & 1 & 16. & 8.9 & 0.32 &
     5.62 \\ \hline G & 30 & 0.1 & 3 & 300 & 1 & 15. & 4.1 & 0.70 &
     2.2 \\ \hline H & 30 & 1.0 & 3 & 30 & 1 & 16. & 4.9 & 0.58 & 3.6
     \\ \hline \hline
   \end{tabular}
\caption{Parameters of numerical models for the chamber problem. 
$\mean{\sigma}_{\rm f}$ is
the final overall magnetization of plasma, $\eta_{\rm r,max}$ is the
maximum radiation efficiency given by Equation~\ref{e10}, and $t_{50}$
is the time by which the plasma emits $50\%$ of the energy
corresponding to this efficiency and given in the units of the chamber 
light crossing time. Other parameters are explained in
Section~\ref{sec:chamber}}
\label{tab:models}
\end{table}

\subsection{Numerical simulations}  
\label{sec:ns}

\begin{figure*}
\includegraphics[width=42mm,angle=-90]{figures/x-s1.eps}
\includegraphics[width=42mm,angle=-90]{figures/x-s2.eps}
\includegraphics[width=42mm,angle=-90]{figures/x-s3.eps}
\includegraphics[width=42mm,angle=-90]{figures/x-s5.eps}
\includegraphics[width=42mm,angle=-90]{figures/x-p1.eps}
\includegraphics[width=42mm,angle=-90]{figures/x-p2.eps}
\includegraphics[width=42mm,angle=-90]{figures/x-p3.eps}
\includegraphics[width=42mm,angle=-90]{figures/x-p5.eps}
\includegraphics[width=42mm,angle=-90]{figures/x-u1.eps}
\includegraphics[width=42mm,angle=-90]{figures/x-u2.eps}
\includegraphics[width=42mm,angle=-90]{figures/x-u3.eps}
\includegraphics[width=42mm,angle=-90]{figures/x-u5.eps}
\includegraphics[width=42mm,angle=-90]{figures/x-t1.eps}
\includegraphics[width=42mm,angle=-90]{figures/x-t2.eps}
\includegraphics[width=42mm,angle=-90]{figures/x-t3.eps}
\includegraphics[width=42mm,angle=-90]{figures/x-t5.eps}
\caption{Magnetically driven oscillations in the model X of the 
chamber problem.  From left to
right, the plots show the solution at $t=0.2$, 0.42, 0.9 and 1.5
respectively. The first row shows the magnetization parameter
$\sigma$, the second row the magnetic pressure $p_m$, the third row
the flow velocity, $u_x=\gamma v_x$, and the bottom row shows the gas
temperature $T=p/\rho$.  }
\label{fig:evolve-x}
\end{figure*}

\begin{figure*}
\includegraphics[width=45mm,angle=-90]{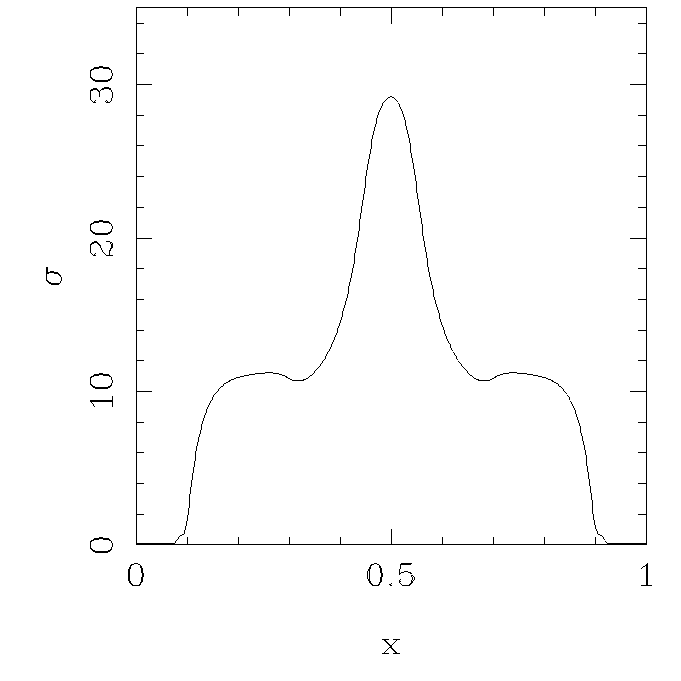}
\includegraphics[width=45mm,angle=-90]{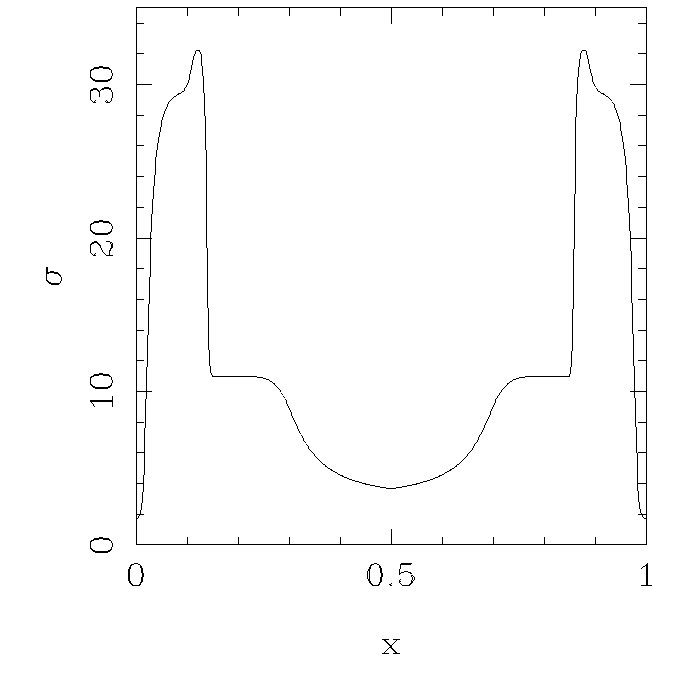}
\includegraphics[width=45mm,angle=-90]{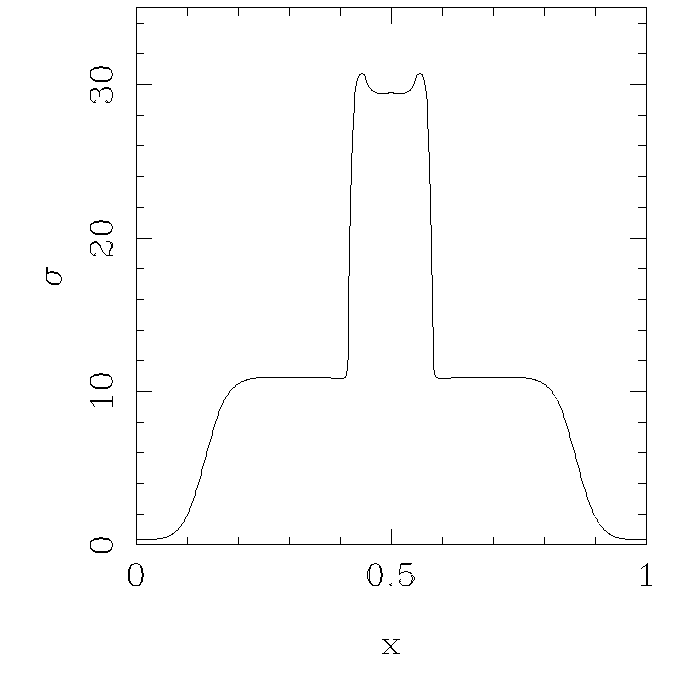}
\includegraphics[width=45mm,angle=-90]{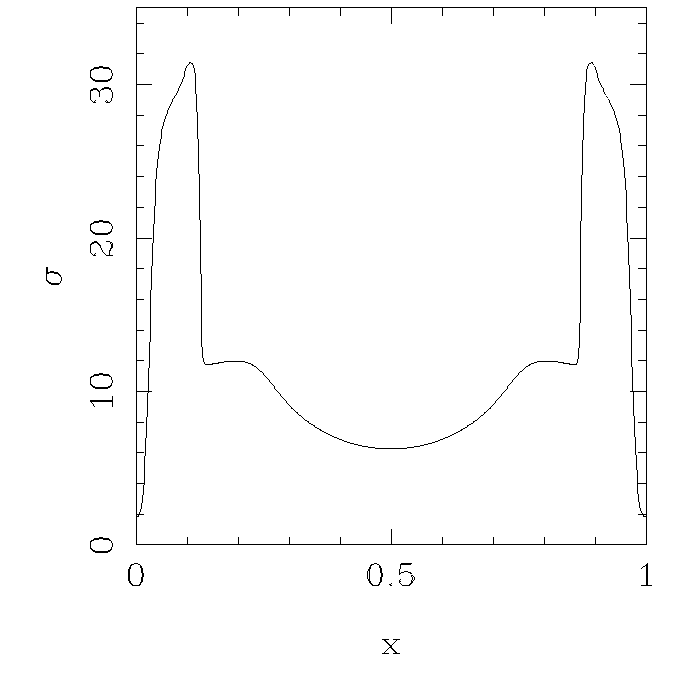}
\includegraphics[width=45mm,angle=-90]{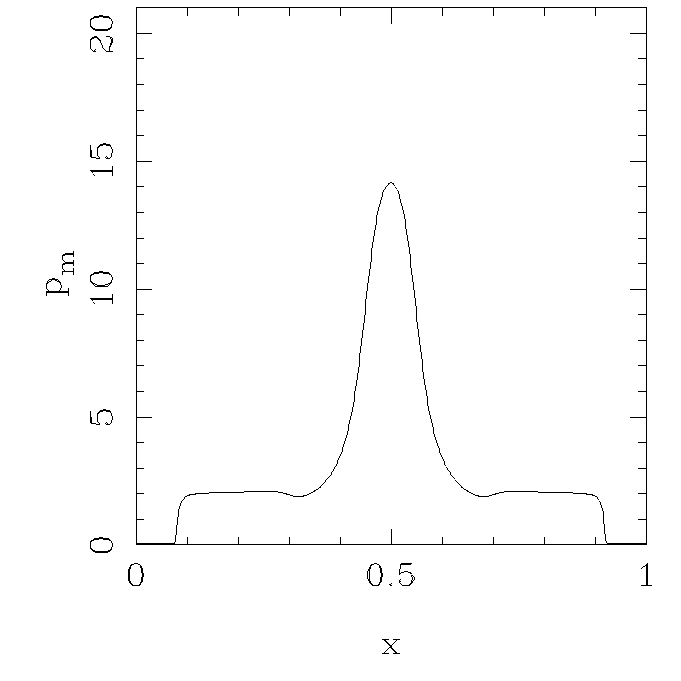}
\includegraphics[width=45mm,angle=-90]{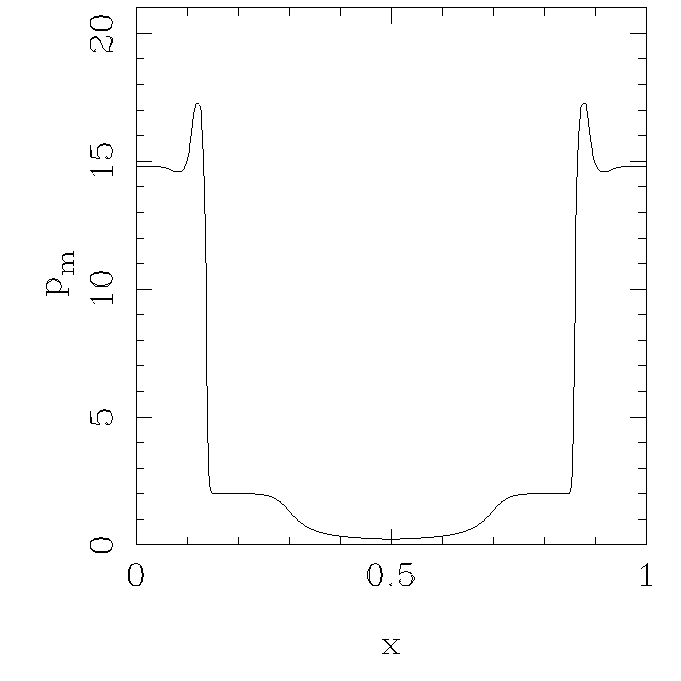}
\includegraphics[width=45mm,angle=-90]{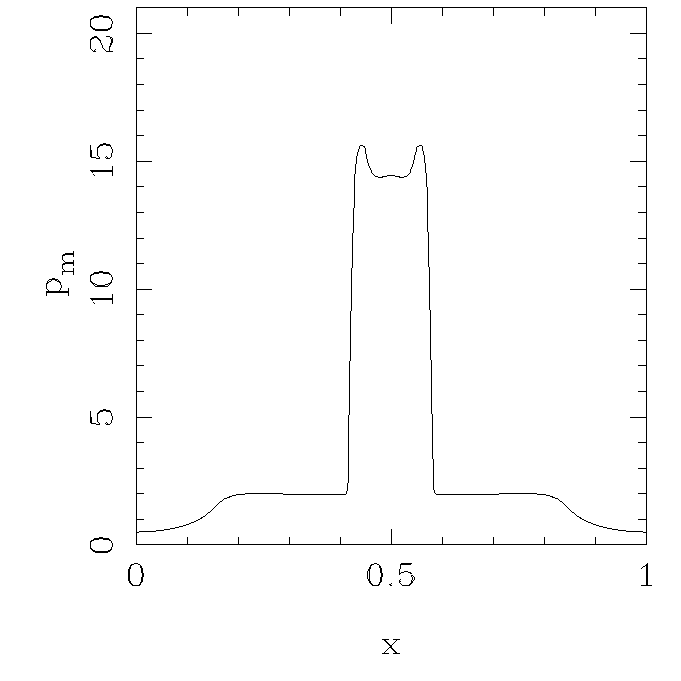}
\includegraphics[width=45mm,angle=-90]{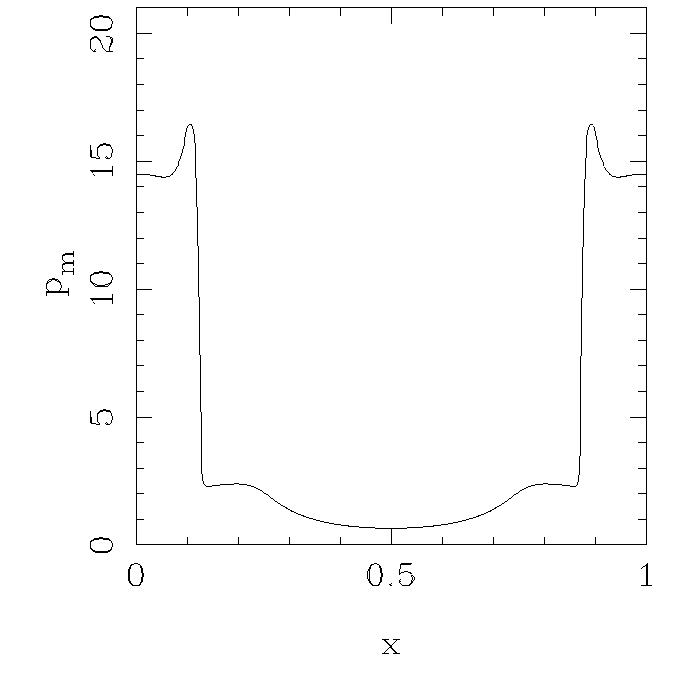}
\includegraphics[width=45mm,angle=-90]{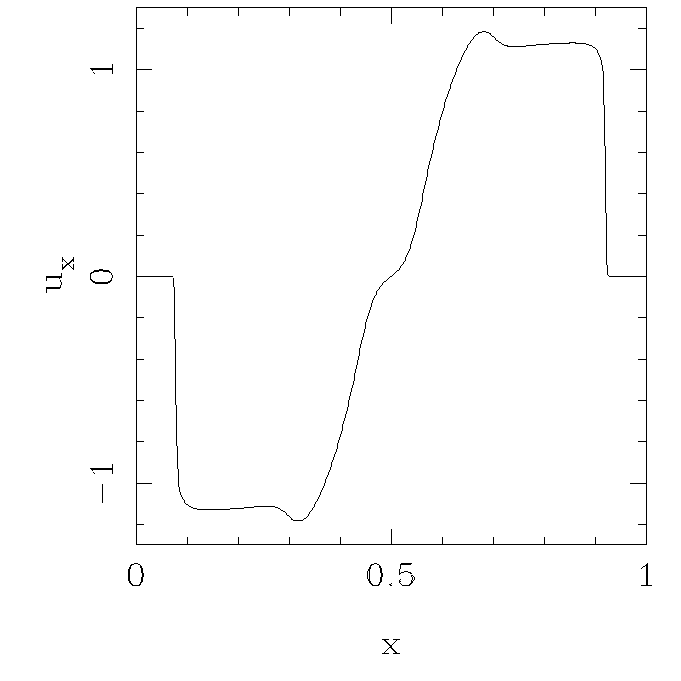}
\includegraphics[width=45mm,angle=-90]{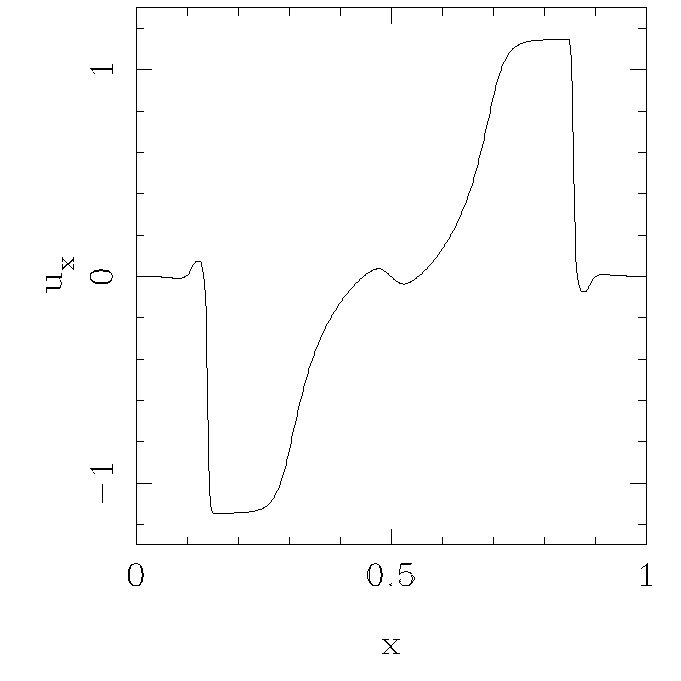}
\includegraphics[width=45mm,angle=-90]{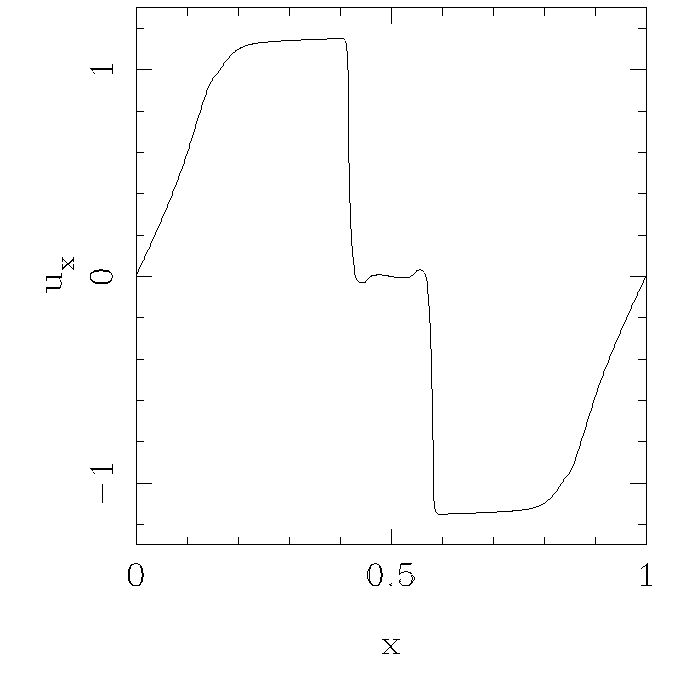}
\includegraphics[width=45mm,angle=-90]{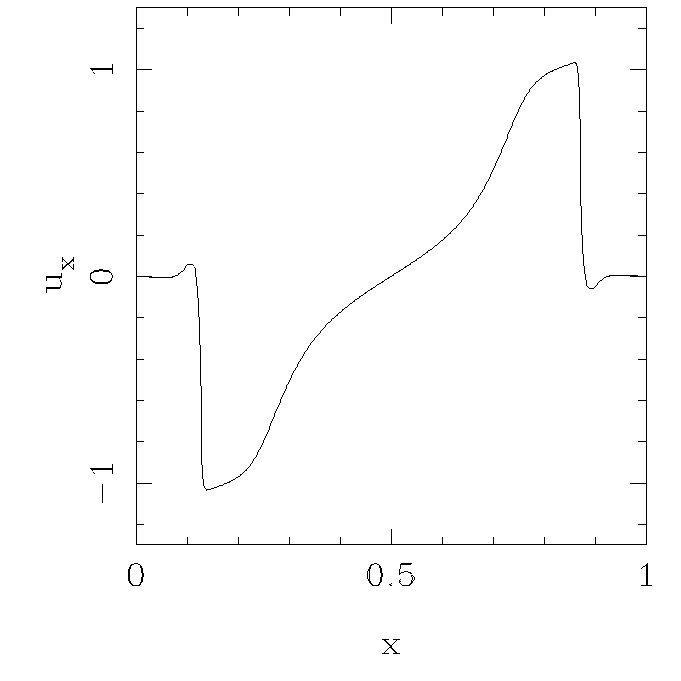}
\includegraphics[width=45mm,angle=-90]{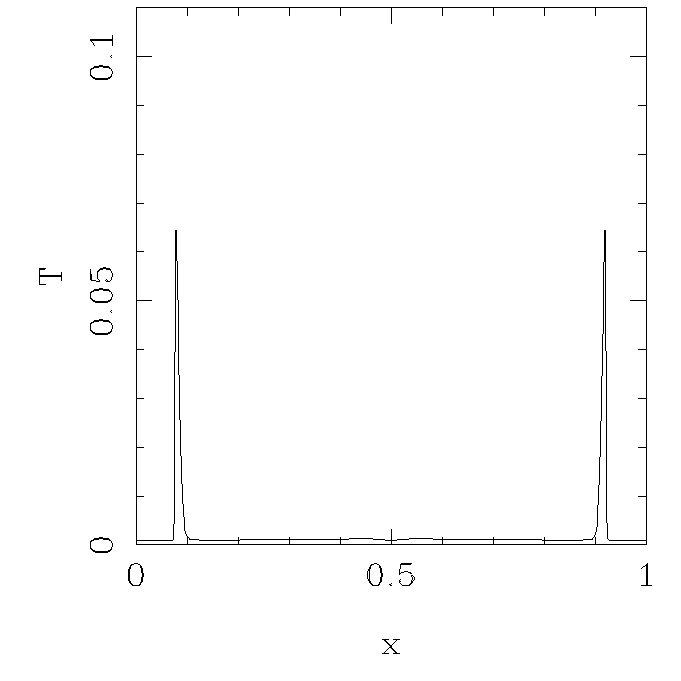}
\includegraphics[width=45mm,angle=-90]{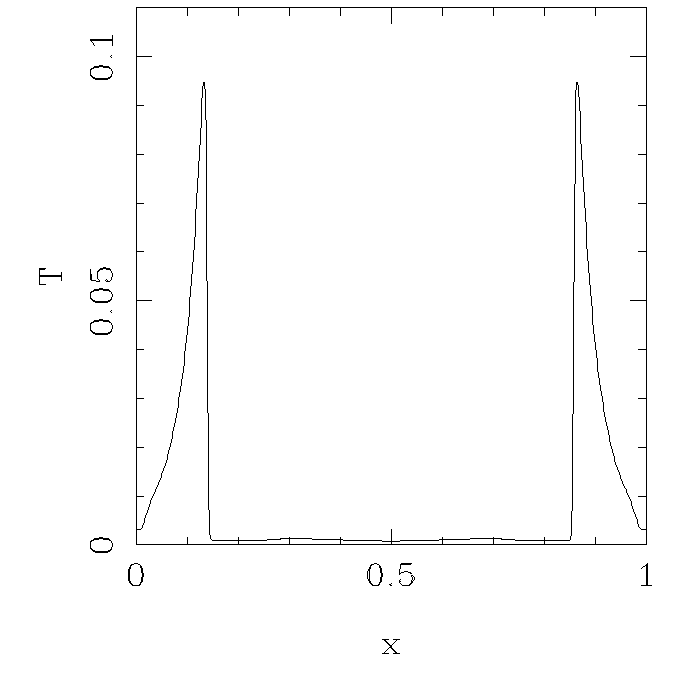}
\includegraphics[width=45mm,angle=-90]{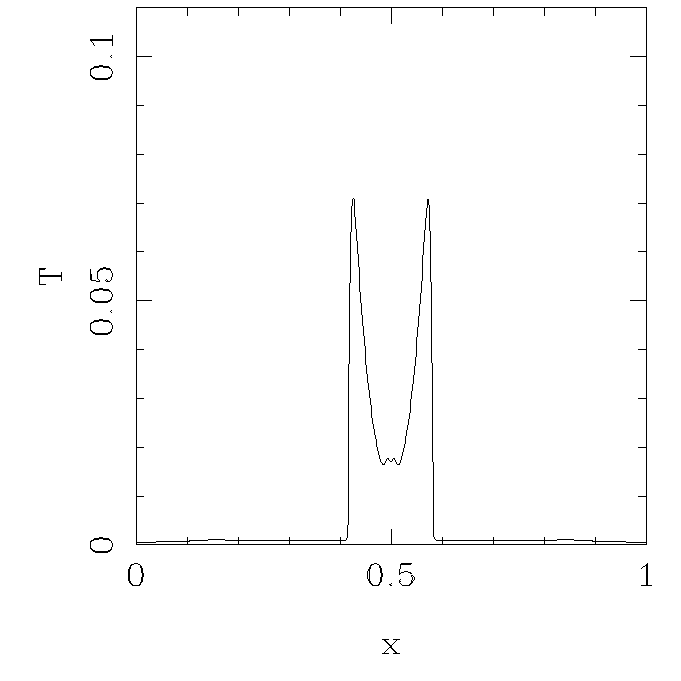}
\includegraphics[width=45mm,angle=-90]{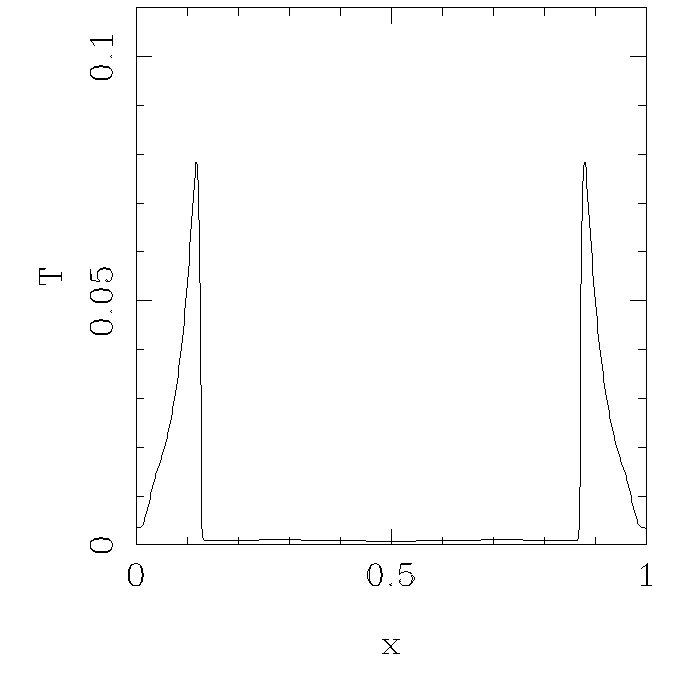}
\caption{Magnetically driven oscillations in the model D of the chamber problem.  
From left to right, the plots show the solution at $t=0.2$, 0.42, 0.9 and 1.5
respectively. The first row shows the magnetization parameter
$\sigma$, the second row the magnetic pressure $p_m$, the third row
the flow velocity, $u_x=\gamma v_x$, and the bottom row shows the gas
temperature $T=p/\rho$.  }
\label{fig:evolve}
\end{figure*}

We solve numerically the one dimensional equations of single-fluid
relativistic magnetohydrodynamics in plane (or slab) geometry. In order to
account for the radiative cooling we introduce a source term in the
energy-momentum equation:
\begin{equation}
\nabla_\nu T^{\nu\mu} = q u^\mu,
\label{source}
\end{equation} 
where $T^{\nu\mu}$ is the stress-energy-momentum tensor of fluid,
$u^\mu$ is its 4-velocity, and $q$ is the cooling rate as measured in
the fluid frame. The cooling rate used in our simulations is not based on any 
particular physical mechanism. At this stage, it makes sense only to require 
for the cooling time to be short compared to the dynamical time scale, as this
seems to be required by the observations of GRBs. To be specific, we put 
\begin{equation}
 q = f_c(T)\frac{e'_t}{\Delta \tau_{\rm cool}},
\label{c-rate}
\end{equation}   
where $T=p/\rho c^2$ is the ``temperature''
$$ 
f_c(T)=\left\{
\begin{array}{ll}
       0, & T<T_0\\             
       (T-T_0)/T_0, & T_0<T<2T_0 \\ 
       1, & T>2T_0\\
\end{array}
\right. ,
$$
and $\Delta \tau_{\rm cool}$ is the characteristic cooling time. In the
simulations we used $T_0=10^{-3}$ and $\Delta \tau_{\rm cool} = 0.04 (l/c)$, a small
fraction of the chamber light crossing time.  This cooling function implies
that all particles cool rapidly and this may be rather  
unrealistic, as only relativistic electrons radiate efficiently. 
However, we do not expect this to change the results by more than a factor 
of few because the 'thermal pump' mechanism \citep{KS01} is likely to be 
efficient in re-processing the retained thermal energy. 
The equation of state describes an ideal gas with the adiabatic index
$\Gamma=4/3$.  

The numerical scheme is based on the one described in
\citet{K99}, with few improvements added over the years. The
computational grid is uniform with 600 cells. Because of the presence of 
strong shocks in the solution the scheme is only of the first order 
accuracy. Our convergence study
shows that the typical computational error for most of the parameters 
discussed below is about few percents. 

The units
are selected in such a way that the dimensionless chamber length is
$l=1$ and the speed of light is $c=1$. The pulse is initially located
in the middle of the chamber. The transition from the pulse to the gap
state is smoothed out using the $\tanh(x)$ function, with the width of the
transition layer $\delta x = 0.03$.

Table~\ref{tab:models} shows the parameters of all models we present
here. Following the reasons behind the magnetic pump mechanism, the
magnetization parameter of the pulse, $\sigma_p$, is always high,
varying between 5 and 30. The gap magnetization is normally much
lower, $\sigma_g\sim0.1$, but we also included the cases where $\sigma_g$
is above unity. In all cases, the total energy of the initial state is
dominated by the magnetic field, corresponding to
$\mean{\sigma}_0>1$. 

\begin{figure*}
\includegraphics[width=45mm,height=60mm,angle=-90]{figures/mod-x.eps}
\includegraphics[width=45mm,height=60mm,angle=-90]{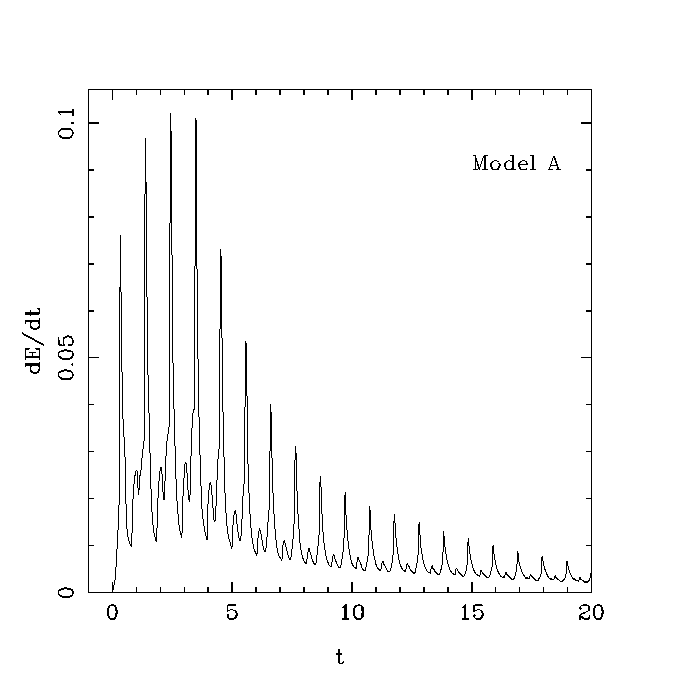}
\includegraphics[width=45mm,height=60mm,angle=-90]{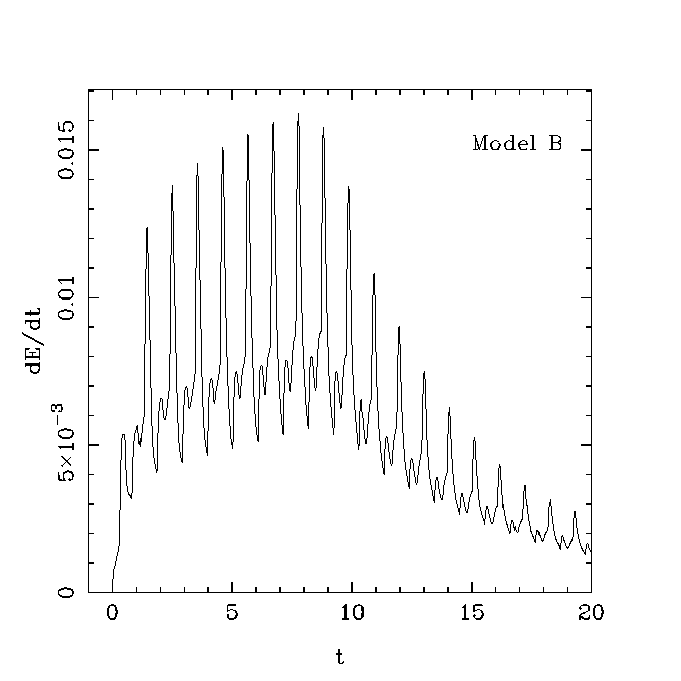}
\includegraphics[width=45mm,height=60mm,angle=-90]{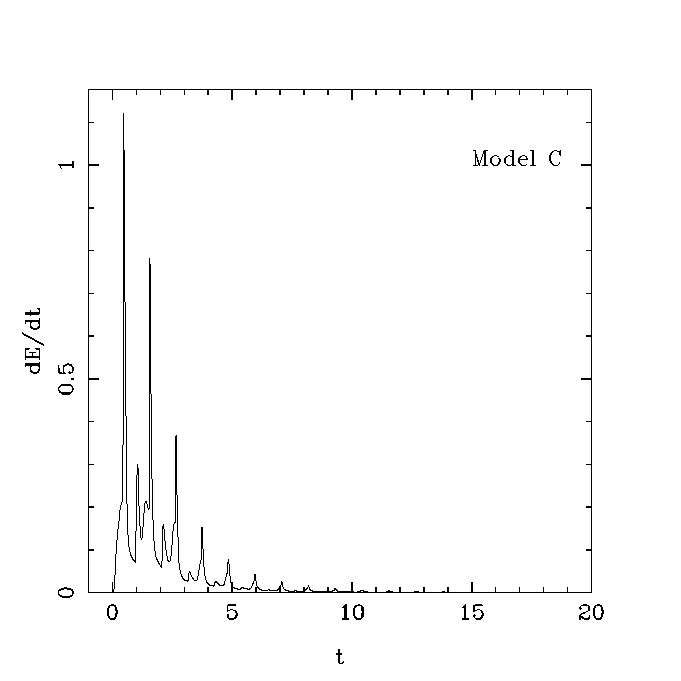}
\includegraphics[width=45mm,height=60mm,angle=-90]{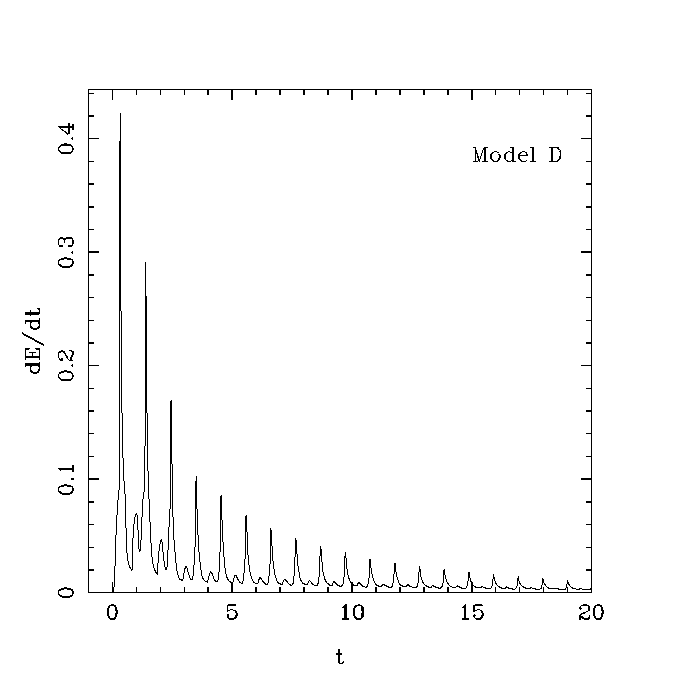}
\includegraphics[width=45mm,height=60mm,angle=-90]{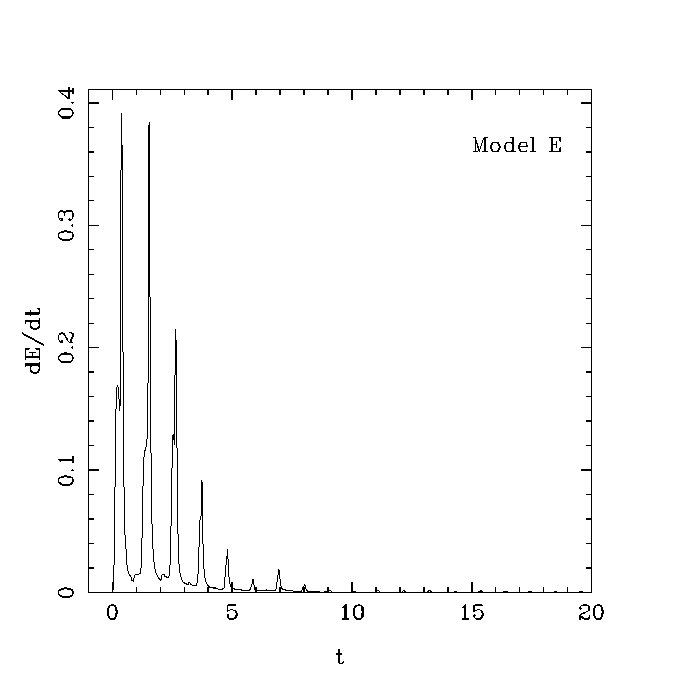}
\includegraphics[width=45mm,height=60mm,angle=-90]{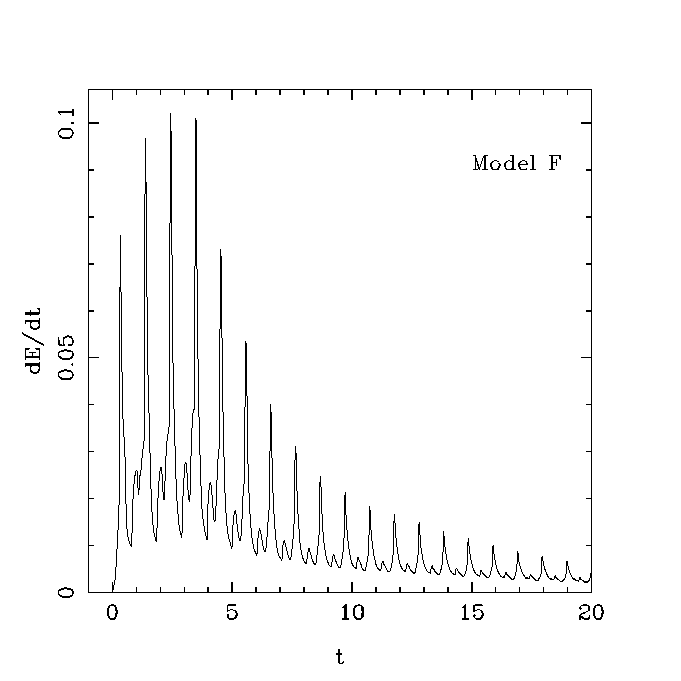}
\includegraphics[width=45mm,height=60mm,angle=-90]{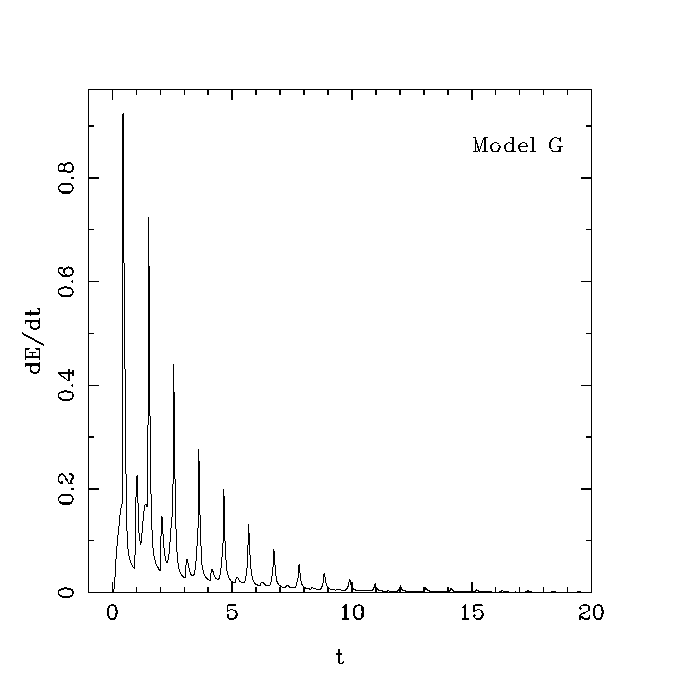}
\caption{ The rate of radiation losses normalized to $E_0/t_0$, where
$E_0$ is the total initial magnetic energy and $t_0=l/c$ is light
crossing time of the chamber.  }
\label{fig:dedt}
\end{figure*}

We first discuss the model X, where the gap is basically empty, as this is  
the closest case to the scenario envisioned in \citet{GKS10}.   
Figure~\ref{fig:evolve-x} illustrates the flow dynamics in this model. 
The first column shows the solution at $t=0.2$, when the
pulse is beginning to expand for the first time. In both halves of the chamber, 
the solution exhibits a rarefaction wave, moving
into the pulse and ejecting its plasma into the gap. The ejected plasma expands 
into the gap space and develops very high velocity. In fact, it can reach  
$\gamma=2\sigma_0+1$ when the gap is a pure vacuum \citep{L10,GKS10}, but in 
our simulation it is limited by the numerical resolution and the non-vanishing 
density of the gap plasma. 

The second column shows the solution at $t=0.42$, soon after the
collision of the flow with the chamber walls, which drives strong shock 
waves back into the pulse. One can see that behind the shocks both 
the magnetic pressure and the plasma magnetization are almost as high as initially
in the pulse. The gas passed through the shock is heated to a very high 
temperature and its cooling rate reaches maximum. On the contrary, in the initial 
location of the pulse a secondary gap is forming behind the two rarefaction 
waves which have now been reflected off the centre. The gap and the pulse have 
switched their places almost recreating the initial conditions.  

The third column shows the solution at $t=0.9$, soon after the
reflected shocks have collided in the center of the chamber for the first time.
One can see that a region of high magnetic pressure and $\sigma$ is formed in
the original location of the pulse. This ``born again'' pulse drives
the next round of expansion and shock heating.   

The forth column of Figure~(\ref{fig:evolve-x}) shows the solution at $t=1.2$, 
which is separated from the solution presented in the first column by 
exactly one light crossing time of the chamber. Qualitatively, the solutions look 
very similar, but at $t=1.2$ the gap is no longer empty and the pulse expansion 
drives a shock through its plasma. This is why the Lorentz factor is so much 
lower. 

Figure~\ref{fig:evolve} shows for comparison the solution for the  
model D, where the gap is filled with a significant amount of plasma from 
the start.  One can see that there is a strong similarity between models
D and X. The first column shows the solution at $t=0.2$, when the
pulse is beginning to expand and compress the gap plasma for the first
time. In both halves of the chamber the solution exhibits a rarefaction wave, moving
into the pulse, and a shock wave, moving into the gap. At the shock, 
the plasma is heated and then it rapidly cools. 
An insignificant heating is also seen in the middle of the
rarefaction wave, where the velocity gradient is the highest.  This
heating is entirely due to the numerical viscosity, and represents 
computational errors. The second column shows the solution at $t=0.42$, soon 
after the shocks reflection off the chamber walls. By this
point the gap plasma has been shocked twice, first by the incident and then 
by the reflected shock. It's temperature and the cooling rate are reaching 
maximum. By $t=0.9$, the reflected shocks have collided in the center of the 
chamber for the first time and the ``born again'' pulse drives
the next round of expansion and shock heating.  When
the shocks reflect of the walls again (see the fourth column of
Figure~\ref{fig:evolve}) the solution is very similar to that at the
time of the first reflection. Thus, the system is undergoing
strong oscillations and the gap plasma experiences repeated shock-heating.

Figure~\ref{fig:dedt} shows the integral rate of radiative cooling  for all 
our models, except G.  These curves exhibit strong peaks at the time 
of the shock collisions at the walls and in the middle of the chamber, 
as this is the time when the shocks cross plasma with lower magnetization. 
In model X, the amplitude of the peaks decreases with time 
monotonically. This reflects the fact that in this models the gaps developing 
in the centre of the chamber have very similar parameters to those developing 
at the walls. The overall decrease of the cooling rate is caused by 
the gradual decrease of the shock strength and increase of the gap 
magnetisation. 
In all other models, peaks corresponding to the shock collision with the wall
are significantly stronger than those corresponding to the 
shock collisions in the middle of the chamber. This is because the  
gaps are filled with weakly magnetised plasma from the start. 
 
These peaks are strongest for models C,D,E and G,
which have lower gap magnetization, $\sigma_g=0.1$, and much weaker in
models A,B,F, and H which have higher gap magnetization, $\sigma_g=1,3$.
The integral radiative cooling rate of models with higher gap magnetization
is also much lower and it declines slower. This is fully consistent
with lower energy dissipation rate in the case of higher magnetization, as
explained in Section~\ref{sec:shock-eff}, and with the fact that the overall
radiative efficiency does not depend on the magnetization (see
Section~\ref{sec:chamber}).

\begin{figure}
\includegraphics[width=70mm,angle=-90]{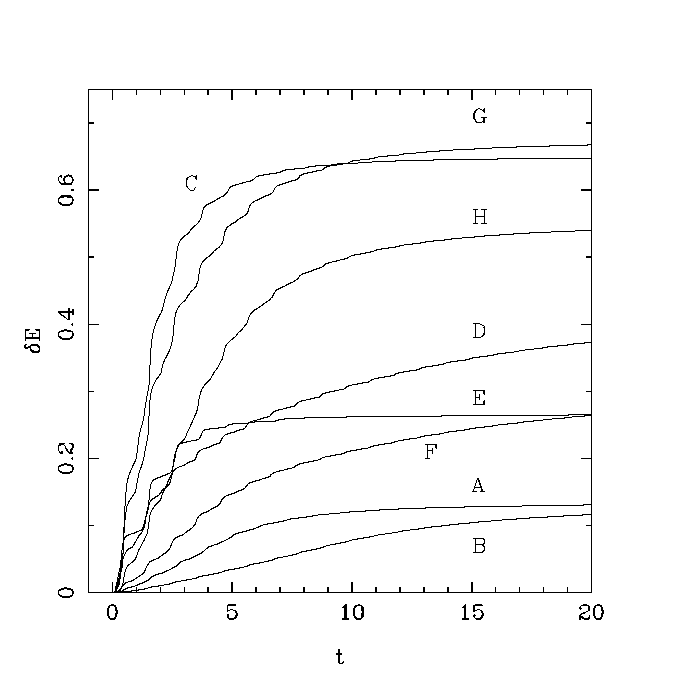}
\caption{Total energy radiated by the time $t$ normalized to the 
initial magnetic energy in the chamber problem. 
The dashed line shows the model X, which initially has an ``empty'' gap.} 
\label{fig:deltae}
\end{figure}

Figure~\ref{fig:deltae} shows the total energy $\delta E(t)/E_0$
radiated by the time $t$. It confirms that even models with
high gap magnetization eventually approach the efficiency $\eta_{\rm
r,max}$ given by Eq.\ref{e10}, this only takes a bit longer.  
The last column of Table~\ref{tab:models} gives the time required for 
the system to radiate 50\% of $\eta_{\rm r,max} E_0$, the total amount of energy 
which will be eventually lost. It spans from one to ten chamber 
light crossing times, increasing as expected with the gap magnetization. 
It is interesting  that the model X is not much different from
models, in spite of the lack of weakly magnetized plasma in the 
initial state.  The explanation for this seems to reside in the ability of 
strong rarefaction waves to supply such plasma and provide it with much 
higher fraction of the kinetic energy compared to the rest of the flow \citep{GKS10}. 
This agrees with 
the fact that faster dissipation is found in the models C and G, 
which have three times wider initial gaps, thus allowing a larger fraction of the ejected 
pulse plasma to develop low $\sigma$ and high $\gamma$.  
The model E also shows faster dissipation. This is due to the fact that its initial gap, 
which is already weakly magnetized ($\sigma_{\rm}=0.1$), is
ten times more massive compared to other models.

\section{Adiabatic flow}
\label{sec:adiabatic}

\begin{figure*}
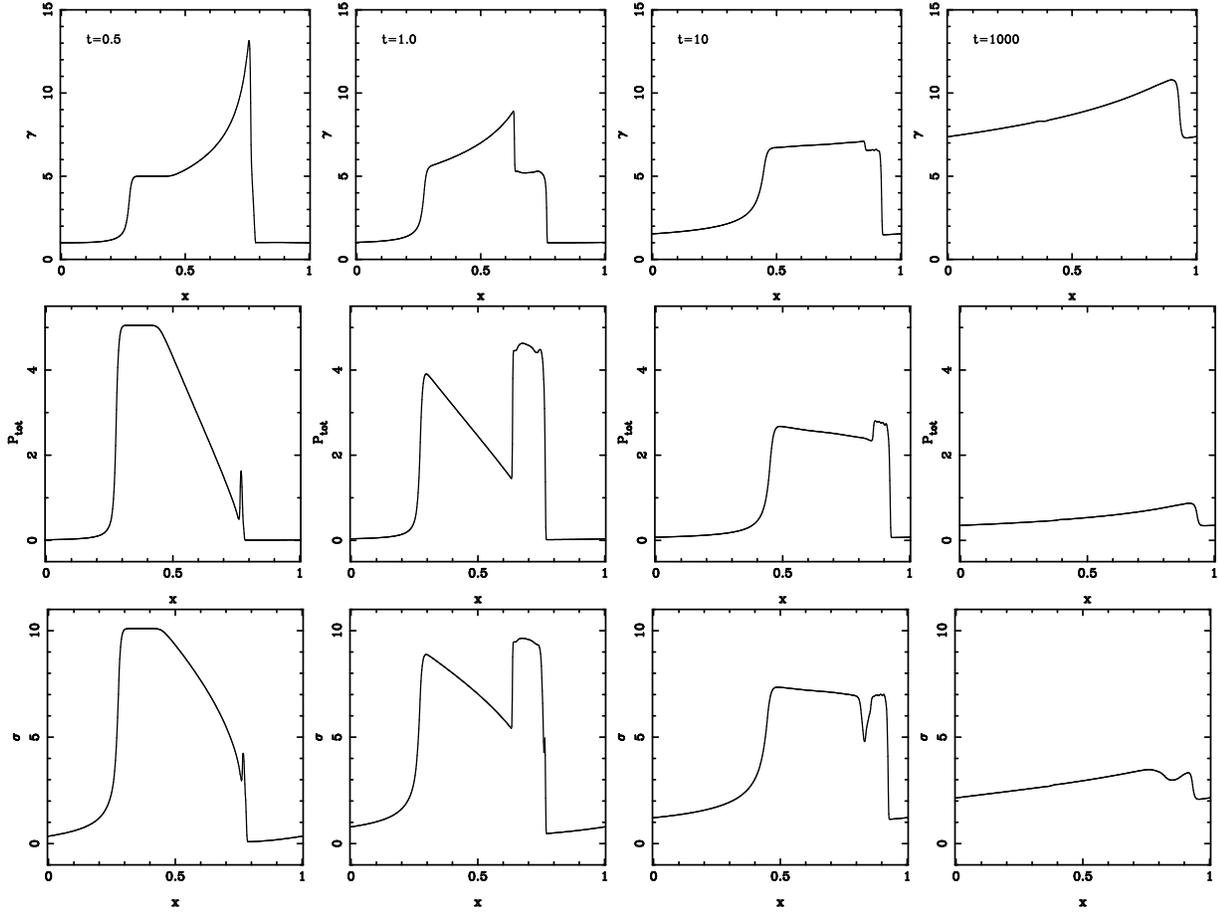

\includegraphics[width=40mm,angle=-90]{figures/ad-lor-0.5.eps}
\includegraphics[width=40mm,angle=-90]{figures/ad-lor-1.0.eps}
\includegraphics[width=40mm,angle=-90]{figures/ad-lor-10.eps}
\includegraphics[width=40mm,angle=-90]{figures/ad-lor-1000.eps}
\includegraphics[width=40mm,angle=-90]{figures/ad-pt-0.5.eps}
\includegraphics[width=40mm,angle=-90]{figures/ad-pt-1.0.eps}
\includegraphics[width=40mm,angle=-90]{figures/ad-pt-10.eps}
\includegraphics[width=40mm,angle=-90]{figures/ad-pt-1000.eps}
\includegraphics[width=40mm,angle=-90]{figures/ad-sig-0.5.eps}
\includegraphics[width=40mm,angle=-90]{figures/ad-sig-1.0.eps}
\includegraphics[width=40mm,angle=-90]{figures/ad-sig-10.eps}
\includegraphics[width=40mm,angle=-90]{figures/ad-sig-1000.eps}
\caption{Various stages in the evolution of the adiabatic flow.  
The first column panels illustrate the solution at $t=0.5$, when the pulse tail 
has just reached its head. This solution is similar to the solution for an isolated 
pulse \citep{GKS10}. The second column shows the solution at 
$t=1.0$. At this point the double shock structure, which has been created by the 
collision of the head and the tail, is already beginning to have a strong impact 
on the flow. The third column shows the solution at $t=10$. Now the flow 
is very different from the isolated pulse solution. The features of strong 
reverse rarefaction wave have been totally erased.     
The last column shows the solution at $t=1000$. By this time the 
flow has developed the ``saw-tooth'' profile characteristic of the classic 
non-linear wave steepening problem. In each column the top plot shows 
the Lorentz factor, the middle plot shows the total pressure, which is dominated by 
the magnetic pressure, and the bottom plot shows the magnetization parameter 
$\sigma$.            
}
\label{fig:adiab-sol}
\end{figure*}

Although suggesting many interesting hints, the chamber problem may differ from the case 
of impulsive flow in many respects. For example, 
in this problem all of the released magnetic energy is converted into radiation. 
This can not be so in the case of a flow, where not only energy but also  
momentum is associated with the electromagnetic field of pulses. Some of this 
momentum can be passed on to plasma, resulting in its bulk acceleration.  
Considering the dynamics of an isolated pulse, \citet{GKS10} have found that the pulse 
develops a very fast ``head'' and that most of the pulse magnetic energy is converted 
into the bulk motion kinetic energy of this head. 
In principle, interaction between pulses may change this outcome. 
Analysing this issue, \citet{GKS10} pointed out that the head does not spread much until  
its acceleration is over and argued that because of this the collisional 
interaction between individual pulses is delayed until the end of their acceleration 
phase. However, each pulse also develops a long slow tail and even if the gaps between 
pulses are completely evacuated initially they quickly become filled with the plasma 
of these tails. The shock interaction between the heads and the tails may modify the 
flow dynamics. 

The easiest way to address this issue is to consider a periodic train of identical 
traveling pulses separated by empty gaps. As the pulses interact and the shocks 
dissipate the kinetic energy of relative 
motion, the flow gradually approaches a state with uniform total pressure.     
Provided the radiative cooling is negligible, the parameters of this state can be found
analytically from the equations of mass, magnetic flux, energy, and momentum conservation.  
 
\subsection{Asymptotic state } 
\label{sec:fss1}

These conservation laws are  
\begin{equation}
 [\rho\gamma l] = 0 \, ,
\label{a1}
\end{equation}   
\begin{equation}
 [B l] = 0 \, ,
\label{a2}
\end{equation}   
\begin{equation}
 [(w\gamma^2 -p +B^2(1-1/2\gamma^2))l] = 0 \, ,
\label{a3}
\end{equation}   
\begin{equation}
 [(w\gamma^2 +B^2)vl] = 0 \, ,
\label{a4}
\end{equation}   
where $w=\rho+\kappa p$ is the relativistic enthalpy, 
$\kappa=\Gamma/(\Gamma-1)$, where $\Gamma$ is the adiabatic index,
$B$ and $v$ are the magnetic field and the plasma speed as measured in 
the observer frame. 
Here, the expressions of the type $[A]$ stand for the difference between the final 
and the initial value of $A$. That is $[A]=A_1-A_0$, where the index ``0'' 
corresponds to the initial state and the index ``1'' to the final state.    
For the initial state $l_0=l_{\rm p}$ is the pulse width, 
whereas for the final state $l_1=l_{\rm p}+l_{\rm g}$ is the wavelength of 
this periodic configuration, as measured in the observer frame. 
        
Denoting the cell mass and magnetic flux as $M=\rho\gamma l$ and $\Psi=Bl$, 
the equations of energy and momentum can be written as
\begin{equation}
 M[\gamma] +[pl(\kappa\gamma^2-1)]  +\Psi^2[\frac{1}{l}(1-1/2\gamma^2)] = 0 \, ,
\label{a3a}
\end{equation}
\begin{equation}
 M[\gamma v] +[pl\kappa v\gamma^2]  +\Psi^2[\frac{v}{l}] = 0 \, .
\label{a4a}
\end{equation}
These equations are to be solved for the thermodynamic pressure, $p_1$, and 
the Lorentz factor, $\gamma_1$, of the final state. Subtracting them and using the 
approximation $v\simeq 1 -1/2\gamma^2$, one finds 
\begin{equation}
 M\left[\frac{1}{\gamma}\right] = (2-\kappa)[pl] \, , 
\label{a5}
\end{equation}
Since $p_0=0$ this yields 
\begin{equation}
 p_1 = \frac{M}{l_1(2-\kappa)}\left[\frac{1}{\gamma}\right] \, .
\label{a6}
\end{equation} 
Next one can simplify Eq.\ref{a4a} by putting $v=1$, which gives
\begin{equation}
 M[\gamma] +[pl\kappa \gamma^2]  +\Psi^2\left[\frac{1}{l}\right] = 0 \, .
\label{a4b}
\end{equation}
Elimination of $p_1$ from this equation leads to the quadratic equation 
\begin{equation}
 \mu x^2 +(1-\mu)x -(1+\sigma_0 \delta) = 0,
\label{a7}
\end{equation}
where $x=\gamma_1/\gamma_0$, $\mu=\Gamma/(2-\Gamma)$, 
and $\delta=(l_1-l_0)/l_1 = \delta_l/(1+\delta_l)$.  This equation has only one 
physical solution 
\begin{equation}
   x = \frac{\mu-1 + D^{1/2}}{2\mu}, 
\label{a8}
\end{equation}
where $D=(1-\mu)^2 + 4\mu(1+\sigma_0\delta)$. 
It easy to see that $x>1$ and thus the flow is accelerated. The corresponding increase 
of the bulk kinetic energy is 
\begin{equation}
   \frac{ [{\cal E}_{\rm k}] }{{\cal E}_{\rm m,0}} = \frac{1}{\sigma_0} (x-1)\, , 
\label{a9}
\end{equation}
whereas the released thermal energy is  
\begin{equation}
   \frac{ [{\cal E}_{\rm t}] }{{\cal E}_{\rm m,0}} =  \frac{\mu}{\sigma_0} x (x-1)\, , 
\label{a10}
\end{equation}
both quantities being normalised to the initial magnetic energy, ${\cal E}_{\rm m,0}$. 

For $\sigma_0\delta \gg 1$, which implies high initial magnetization and not very 
narrow gaps, one has   
\begin{equation}
   x \simeq \sqrt{\frac{\sigma_0\delta}{\mu}} \gg 1. 
\label{a11}
\end{equation} 
Thus $\gamma_1 \simeq \gamma_0 \sigma_0^{1/2}$, which is significantly lower 
compared to $\gamma_1=\gamma_0 \sigma_0$, corresponding to the total 
conversion of magnetic energy into the kinetic one. Hence  
most of the magnetic energy must be converted into heat. 
Indeed, from Eqs.(\ref{a9}) and (\ref{a10}) one has 
\begin{equation}
   [{\cal E}_{\rm t}]  \simeq \frac{x}{\mu} [{\cal E}_{\rm k}] \gg [{\cal E}_{\rm k}] \, . 
\label{a12}
\end{equation}

\begin{figure}
\includegraphics[width=80mm,angle=-90]{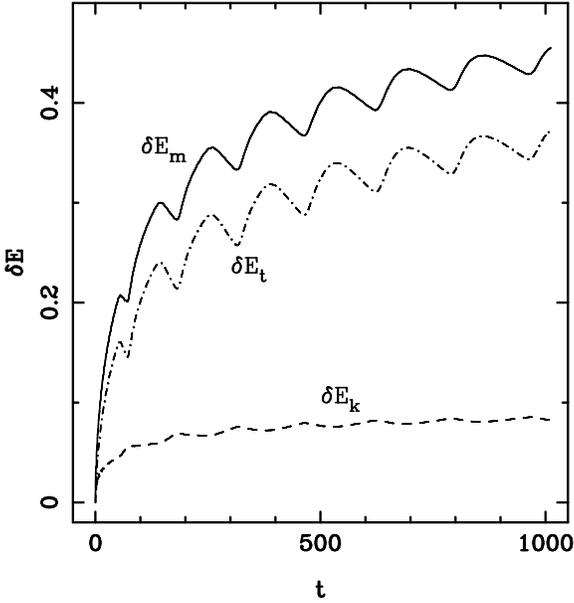}
\caption{Energy balance of the adiabatic flow.  
The dot-dashed line shows the fraction of initial magnetic energy turned 
into heat, the dashed line shows 
the fraction of initial magnetic energy converted into the bulk motion kinetic 
energy, and the solid line shows the total fraction of utilised magnetic energy. }
\label{fig:ad-energy}
\end{figure}

\subsection{Numerical simulations}  
\label{sec:ns1}

In these simulations, we utilise the spatial grid which moves relative to the inertial 
frame of our fiducial observer with the initial speed of the pulse. 
This is similar to the so-called ``moving window'' approach. Namely, we 
use the time-like foliation of space-time defined by the time of observer's  
inertial frame, but introduce new spatial coordinate via the transformation 
$x\to x-\beta_0 t$. This leads to the metric form
\begin{equation}
    ds^2=(-1+\beta_0^2)dt^2+2\beta_0 dxdt +dx^2 +dy^2 +dz^2\, .
\label{a16}
\end{equation}

The computational domain covers one wavelength of the flow and the 
initial solution describes  
a uniform pulse located right in the middle of this domain. The gaps are not absolutely 
empty but they are so highly rarefied and weakly magnetized that have very 
little effect on the solution. The initial Lorentz factor is uniform throughout the
domain. Both at the left and the right boundaries we impose the periodic boundary 
conditions. 

Figure \ref{fig:adiab-sol} illustrates the typical evolution of such a flow.  
In this particular model, the initial magnetization of the pulse is
$\sigma_0=10$, its thermal pressure is negligibly small, and it moves with 
the Lorentz factor $\gamma_0=5$ to the right. 
The flow is super fast magnetosonic, with the Mach number $M\simeq 1.6$. 
The pulse ($0.25<x<0.75$) and the gap are 
equal in linear size, with $l_{\rm p}=l_{\rm g}=0.5$.  The ratio of specific 
heat is again $\Gamma=4/3$ ( $\kappa =4$ and $\mu=2$). 

At $t=0.5$ the solution is very much as this was anticipated in \citet{GKS10}.  
Indeed, it is dominated by two strong rarefaction waves, both originated at the 
pulse boundaries and moving inside the pulse. However, the one produced at the head 
propagates through the pulse much faster than the one produced at the tail. 
As the result, inside most of the pulse the magnetic pressure declines towards the 
head and the magnetic pressure force accelerates the flow.  
The other rarefaction gets stuck at the back where it ejects pulse plasma into 
the tail, which grows in size almost at the speed of light. 
By $t=0.5$ the tip of the tail has already crossed the gap and collided with 
head (of the other pulse). Two shock waves, one forward and one reversed, are produced 
as result of this collision, they are responsible for the spike observed
at $x\simeq 0.77$.

\begin{figure*}
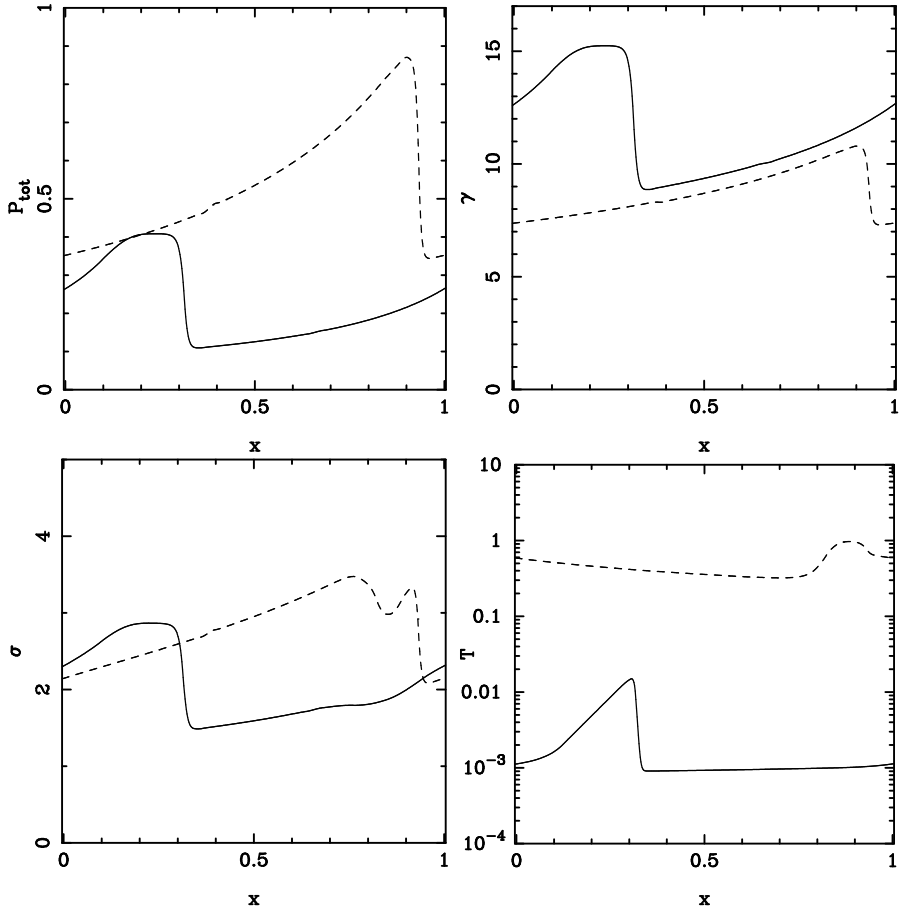

\includegraphics[width=60mm,angle=-90]{figures/rad-pt-1000.eps}
\includegraphics[width=60mm,angle=-90]{figures/rad-lor-1000.eps}
\includegraphics[width=60mm,angle=-90]{figures/rad-sig-1000.eps}
\includegraphics[width=60mm,angle=-90]{figures/rad-t-1000.eps}
\caption{Differences between the adiabatic flow and the cooling flow 
with the same initial parameters ($\sigma_0=10$, $l_{\rm p}=l_{\rm g}=0.5$, 
$\Gamma=4/3$, and  $\gamma_0=5$) at the time $t=1000$. Solid lines show 
the cooling flow and dashed lines the adiabatic one.  
}
\label{fig:ad-rad-sol}
\end{figure*}

At $t=1$ the solution already looks rather different from that of an isolated 
pulse. The reverse shock has moved well inside the pulse and it is now located 
at $x\simeq 0.63$. Behind the shock, the flow has almost recovered its initial Lorentz 
factor, magnetization $\sigma$, and magnetic pressure. 

By $t=10$ all the features of the strong rarefaction wave, characteristic for the 
isolated shell solution, have gone. Now the solution can be broadly described
as a flat-top pulse with a tail.      

Due to the periodic boundary conditions, when a wave reaches one boundary it 
re-appears from the other one. 
The forward and reverse waves do this at a very different rate. 
Indeed, if the wave Lorentz factor in the flow frame is $\gamma_{\rm w} \gg 1$ and 
the flow Lorentz factor $\gamma\gg 1$ then the observed relative speed of the 
forward wave is 
\begin{equation}
  \delta\beta_{\rm fw} = |\beta_{\rm fw}-\beta| \simeq \frac{1}{2\gamma^2} \, ,
\label{beta-fw}
\end{equation}
whereas for the reverse wave we have 
\begin{equation}
  \delta\beta_{\rm rw} = |\beta_{\rm rw}-\beta| \simeq 
   \frac{2\gamma_{\rm w}^2}{\gamma^2 +\gamma_{\rm w}^2} \, ,
\label{beta-rw}
\end{equation}
which is much faster.  
This difference must be behind the observed much more rapid decay of reverse 
waves compared to forward waves. As one can see in Figure \ref{fig:adiab-sol},  at 
$t=1000$ the amplitude of forward waves is much higher and the flow exhibits  
the characteristic ``saw-tooth'' profile similar to the one which develops 
in the problem of nonlinear wave steepening.       
 
Figure \ref{fig:ad-energy} illustrates the energy evolution of the flow. 
It shows the variations of the magnetic, bulk motion, and thermal energies 
per train wavelength, all being normalised to the initial magnetic energy.
Since, in this problem $\delta=0.5$, by the time the system relaxes to a uniform 
state its magnetic energy decreases by 50 percent. As one can see in 
Figure \ref{fig:ad-energy}, this has almost been achieved at $t=1000$. 
According to the analysis of Sec.\ref{sec:fss1}, the asymptotic parameters of  
this flow are  $\gamma\simeq2\gamma_0=10$, $\sigma\simeq 2.5$ and 
$\Delta {\cal E}_{\rm k} \sim 0.1$. These are indeed very close to the 
values observed at $t=1000$ (see Figures~\ref{fig:ad-energy}~and~\ref{fig:adiab-sol}).

The characteristic time of relaxation towards the uniform state must be given by   
the dissipation time scale of the forward waves associated with the saw-tooth 
structure of the flow. First, the pulse has to pass through the forward fast shock 
where the dissipation occurs. Given the result (\ref{beta-fw}), the corresponding 
time, $\Delta t_{\rm c} \simeq 2l\gamma^2$, is actually independent on the shock 
speed, and hence its strength. Second, each time when a strong shock crosses the 
pulse it dissipates only a fraction $\sim 1/2\sigma$ of the available pulse energy 
(see Eq.\ref{eff}). Thus, the relaxation time can be estimated as 
\begin{equation}
\Delta t_{\rm r} \simeq 4 l\sigma\gamma^2\, .
\label{t_relax}
\end{equation} 
This turns out to be about the same as the time scale of the nonlinear steepening of 
a fast magnetosonic wave \citep[see Eq.11 in][]{L03}. For our numerical model, 
$\Delta t_{\rm r} \simeq 1000$, if we use the parameters of the initial solution for 
the calculations, in excellent agreement with the numerical results.

\subsection{Comments on the geometrical effects}

   The results of our study show that the interaction between individual pulses 
(magnetic shells) becomes important well before the coasting phase of an isolated 
pulse, significantly reducing the efficiency of impulsive acceleration. 
However, the dynamics of adiabatic flows is strongly influenced by their 
geometry. For example, the sideways 
expansion of conical jets is an efficient way of converting their thermal 
energy into the kinetic energy of bulk motion.  Thus, provided the radiative cooling 
time is long compared to the adiabatic one, a conical flow can still eventually become 
kinetic energy dominated. In the context of the pulsar wind acceleration 
this mechanism has been discussed in \citet{L03}.   

\begin{figure*}
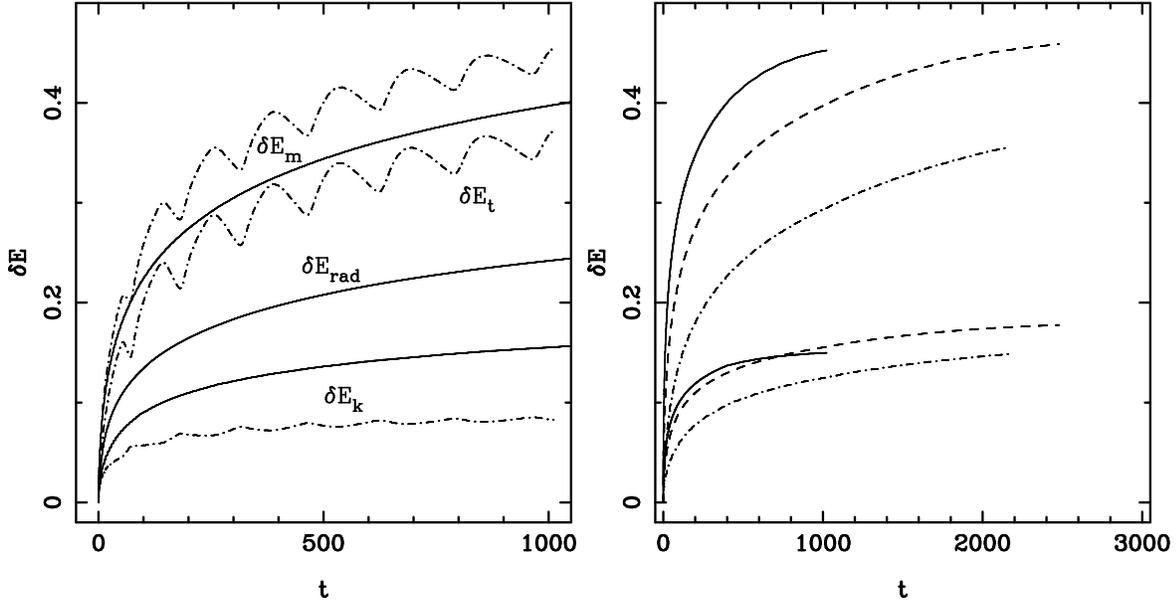

\includegraphics[width=80mm,angle=-90]{figures/ad-rad.eps}
\includegraphics[width=80mm,angle=-90]{figures/rad-energy.eps}
\caption{{\it Left panel}: Energy balance of the adiabatic and  
the radiatively cooling flows with the same initial parameters. The solid lines show 
the total fraction of utilized magnetic energy (the top line), the fraction of
the magnetic energy converted into radiation (the middle line), and the 
fraction of the magnetic energy converted into the bulk motion kinetic
energy (the bottom line) for the model with fast radiative cooling. 
The dash-dotted lines show the total fraction of utilized magnetic 
energy (the top line), the fraction of the magnetic energy converted into 
heat (the middle line), and the
fraction of the magnetic energy converted into the bulk motion kinetic
energy (the bottom line) for the adiabatic model.
{\it Right panel}: Energy balance of fast cooling flows with different initial 
Lorentz factors. There are three pairs of lines, 
solid lines for $\gamma_0=3$, 
dashed lines for $\gamma_0=5$, 
and dash-dotted lines for $\gamma_0=10$. In each pair, the top line shows 
the total fraction of utilized magnetic energy, and the bottom line shows the 
fraction converted into the bulk motion kinetic energy. The difference between 
the top and the bottom lines gives the fraction of the magnetic energy converted 
into the radiation.   
}
\label{fig:rad-energy}
\end{figure*}

In order to find the asymptotic flow parameters in this case we simply 
notice that because 
\begin{equation}
  {\cal E}_{\rm m,1} =  {\cal E}_{\rm m,0} (1-\delta)  
\end{equation}
the kinetic energy 
\begin{equation}
  {\cal E}_{\rm k,1} =  {\cal E}_{\rm k,0} + {\cal E}_{\rm m,0} \delta \simeq 
  {\cal E}_{\rm m,0} \delta \, .
\end{equation}
Thus, the asymptotic magnetization parameter should be 
\begin{equation}
\sigma_1 \simeq  \frac{{\cal E}_{\rm m,1}}{{\cal E}_{\rm k,1}} = (1-\delta)/\delta \, .
\end{equation}
We comment here that unless the gap is much wider than the pulse, and hence 
$\delta$ is very close to unity, the asymptotic magnetization is still not much lower 
than unity. For example, when $l_p=l_g$, and hence $\delta=1/2$, 
this equation gives $\sigma_1 =1$. Eqs.(\ref{a1}) and (\ref{a2}) give us another 
expression for $\sigma_1$, namely 
\begin{equation}
\sigma_1 = \sigma_0 \frac{\gamma_0}{\gamma_1} (1-\delta) \, .
\end{equation}
Combining the last two equations we find the asymptotic Lorentz factor 
\begin{equation}
\gamma_1 = \gamma_0 \sigma_0\delta \, .
\end{equation}

Although in this case we almost recover the asymptotic parameters of isolated 
shells \citep{GKS10}, the acceleration mechanism is different. Indeed, the key 
role is played by the shock heating and thermal acceleration, instead of the magnetic 
pressure acceleration.  Moreover, in the context of GRBs,   
this regime is not very attractive  because it implies low radiative efficiency. 
Indeed, by the time the flow becomes kinetic energy dominated, its impulsive origin is 
``washed out'', with the remaining shock waves being week and allowing dissipation 
of only a small fraction of the flow power.

\section{Radiatively cooling flow}
\label{sec:FCF}
 
On the opposite extreme to the adiabatic flow is the case where the 
radiative cooling time scale is much shorter compared to the time scale 
of adiabatic cooling. This is 
the so-called ``fast radiative cooling'' regime, which has been often 
discussed in connection to GRB jets. In this case the difference between the plane 
and spherical geometry is unimportant as all of the released 
thermal energy is lost to the radiation. Assuming that the total fraction of 
the utilised magnetic energy is the same as in the adiabatic regime,  
one would expect very high radiative efficiency,  
\begin{equation}
  \eta_{\rm r} = \frac{[{\cal E}_{\rm t}]}{{\cal E}_{\rm m,0}} 
\simeq \delta \, .  
\label{a13}
\end{equation}       
In fact, this is the same as in the chamber problem with empty gaps (see Eq.\ref{e11}). 
However, both the radiative cooling and the radiation reaction 
force may modify the flow dynamics and in this Section we investigate their 
roles.

\begin{figure*}
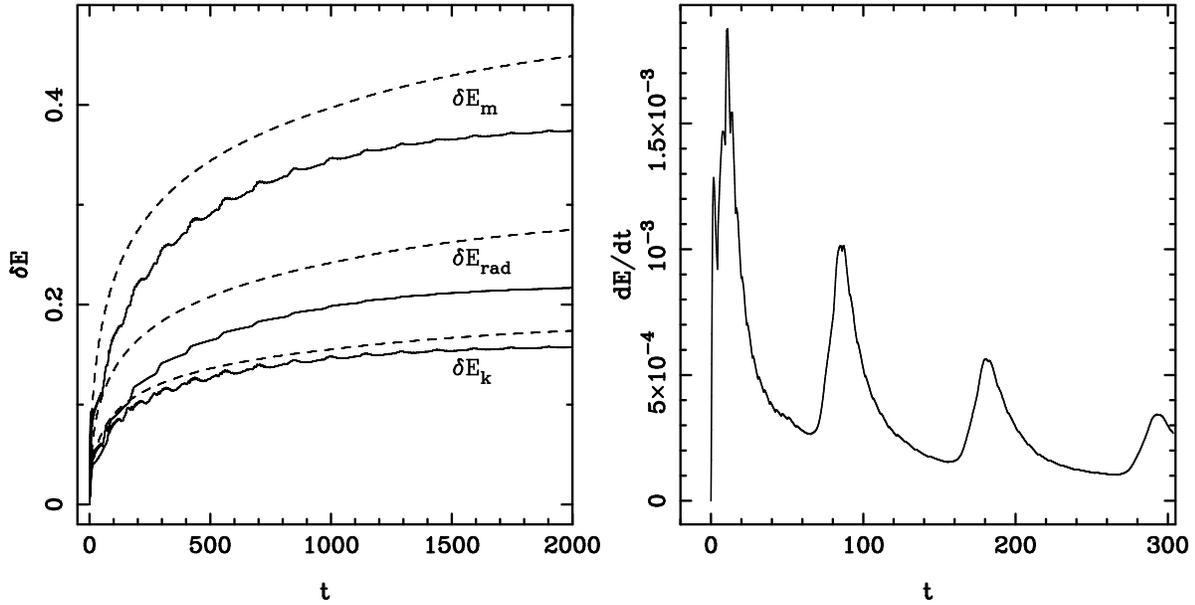

\includegraphics[width=80mm,angle=-90]{figures/mix-vac-comp.eps}
\includegraphics[width=80mm,angle=-90]{figures/rad-loss.eps}
\caption{{\it Left panel}: Energy balance of the radiatively cooling flows 
with initially empty gaps (dashed lines) and gaps filled with low-magnetized 
plasma ($\sigma=0.1$) of the same mass (solid lines). 
{\it Right panel:} Radiative energy loss rate for the model with filled gaps.
}
\label{fig:mix}
\end{figure*}

\subsection{Asymptotic state} 
\label{sec:fss2}

In this case, one can still try to determine the asymptotic flow parameters
using the conservation laws, just like this was done in Sec.\ref{sec:adiabatic}. 
While the mass and the magnetic flux conservation laws remain the same, the 
laws for energy and momentum have to be modified in order to account for the radiative 
losses: 
\begin{equation}
 [(\rho\gamma^2 +B^2(1-1/2\gamma^2))l] = Q \gamma_{\rm rad} \, ,
\label{b3}
\end{equation}
\begin{equation}
 [(\rho\gamma^2 +B^2)vl] = Q \gamma_{\rm rad} v_{\rm rad} \, . 
\label{b4}
\end{equation}   
Here we use the fact that the total energy-momentum of emitted photons 
can be written as $Q u^\nu_{\rm rad}$ (see Eq.\ref{source}), 
where $u^\nu_{\rm rad}$ is the averaged 4-velocity of the flow during its 
relaxation. Its value depends on possible correlation between 
the rate of radiative losses and the flow speed, e.g. the radiation may come 
mostly from the fastest parts of the flow.    
Since this is essentially an unknown parameter, we end up with 
an under-determined system, which has only four equations and five unknowns, 
namely $\gamma_1$, $B_1$, $\rho_1$, $\gamma_{\rm rad}$, and $Q$. 

Eliminating $Q$, $B_1$, $\rho_1$, and using the usual high speed 
approximation $v \simeq 1-1/2\gamma^2$, we find the equation determining 
$\gamma_1$  as a function of  $\gamma_{\rm rad}$,
\begin{equation}
\left(\frac{\gamma_1}{\gamma_0}\right) 
\left(1-\frac{\gamma_{\rm rad}^2}{\gamma_1^2}\right) - 
\left(1-\frac{\gamma_{\rm rad}^2}{\gamma_0^2}\right) = \sigma_0\delta \, .
\label{b5}
\end{equation}
If we put $\gamma_{\rm rad}=\gamma_1$ then this equation yields 
\begin{equation}
  \gamma_1 = \gamma_0(1+\sigma_0\delta)^{1/2}
  \simeq \gamma_0 \sqrt{\sigma_0\delta} \, ,
\label{b6}
\end{equation}  
where the last step assumes $\sigma_0 \delta\gg 1$. This is indeed very similar to what 
we have found in the adiabatic case in the slab geometry (see Eq.\ref{a11}). 

On the other hand, when $\gamma_{\rm rad}=\gamma_0$, we have
\begin{equation}
  \gamma_1 = \frac{\gamma_0}{2}(\sigma_0\delta + (\sigma_0^2\delta^2+4)^{1/2}) 
  \simeq \gamma_0\sigma_0 \delta \, .
\label{b7}
\end{equation} 
This implies very efficient flow acceleration with almost total conversion
of magnetic energy into the bulk motion kinetic energy for $\delta\simeq 1$, 
contrary to what we have anticipated. The reason for this is the strong 
radiation reaction force, which accelerates the flow. Indeed,  
$\gamma_{\rm rad}=\gamma_0$ implies that most of the time  $\gamma_{\rm rad}$ is 
lower compared to the center-of-momentum Lorentz factor. Because of this, 
in the centre-of-momentum frame, the photons are emitted mostly in 
the direction opposite to the flow direction.  This seems hardly possible as 
most of the radiation must come from the immediate downstream of forward shocks 
and if the flow preserves the structure of adiabatic solution then these are 
the fastest sections of the flow (see Figure~\ref{fig:adiab-sol}). In fact, this 
argument suggests that the asymptotic Lorentz factor can be quite close 
to that given by Eq.\ref{b6}.     
In order to verify this deduction we have carried the numerical simulations 
described below.

\subsection{Numerical simulations}  
\label{sec:ns2}

In these simulations we used the same cooling function as in the 
chamber problem (see Eq.\ref{c-rate}). 
In order for the numerical shock structure not to have much 
influence on the overall effect of radiative cooling, we had to insure 
that the cooling length scale was significantly higher compared to the shock 
thickness. If the proper cooling time scale is $\Delta\tau_{\rm cool}$
then in the observer frame this scale is 
$\Delta t_{\rm cool} = \gamma\Delta\tau_{\rm cool}$. In this time,  
a weak forward magnetosonic shock wave moves relative to the flow by 
the observed distance 
$\Delta l \simeq (1/\gamma^2)\Delta t_{\rm cool} \simeq \Delta\tau_{\rm cool}/\gamma$.    
Thus in order for the length of the cooling region behind the shock to be 
a fraction $\alpha=\Delta l/l$ of the flow wavelength, the proper cooling time should 
be $\Delta\tau_{\rm cool} =\alpha\gamma l$. In these simulations we used $\alpha=0.2$.   

First we studied the model with the same initial parameters  
as the adiabatic one described in Section~\ref{sec:adiabatic}:  
$\sigma_0=10$, $l_{\rm p}=l_{\rm g}=0.5$, $\Gamma=4/3$, and  $\gamma_0=5$. 
Overall, the cooling flow shows a similar deviation from the dynamics of isolated 
pulse, considered in \citet{GKS10}, as the adiabatic model. 
Figure~\ref{fig:ad-rad-sol} compares the adiabatic and cooling flows at later times. 
As one can see, both flows develop the characteristic ``saw-tooth'' profiles and apart 
from the temperature, their parameters are rather similar.  The cooling flow is only 
slightly faster compared to the adiabatic one.  
  
\begin{figure}
\includegraphics[width=80mm,angle=-90]{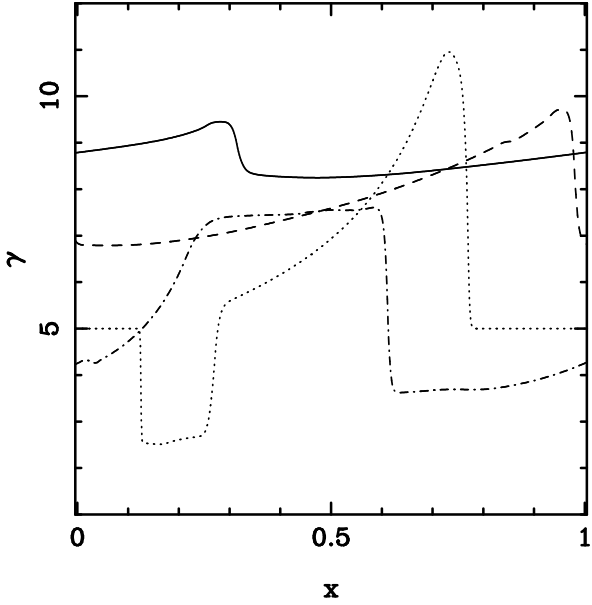}
\caption{ Evolution of the Lorentz factor in the model with filled gaps. 
The curves show the solutions at $t=1$ (dotted line), $t=100$ (dash-dotted line), 
$t=1000$ (dashed line), and $t=3000$ (solid line).   
}
\label{fig:mix-lor}
\end{figure}

The left panel of 
Figure~\ref{fig:rad-energy} compares the energy balance of both flows. It confirms that 
the cooling flow is slightly more efficient in converting the magnetic energy into the 
bulk kinetic energy. It also shows that by $t=1000$ most of the free magnetic energy 
has been utilized. Comparing the numerical results with the predicted asymptotic 
parameters of the cooling flow (see Sec.\ref{sec:fss2}), we find that Eq.\ref{b6} 
does better, with $\gamma_1\simeq 11$, whereas  Eq.\ref{b7} significantly 
overestimates the asymptotic Lorentz factor, giving $\gamma_1\simeq 50$. 
Figure \ref{fig:ad-rad-sol} shows that most of the radiative energy losses are 
associated with the fastest parts of the flow, just like we have anticipated, and 
this explains why Eq.\ref{b6} provides a more accurate estimate. 

The right panel of Figure \ref{fig:rad-energy} compares the energy budgets of three 
cooling models which differ only by the Lorentz factor of the initial solution, 
$\gamma_0=3,\,5,$ and 10. One can see that the efficiency of magnetic acceleration 
grows with $\gamma_0$, but only slightly. In all these models, more than half of the 
free magnetic energy is converted into radiation. Supported by the strong arguments presented 
in Sec.\ref{sec:fss2}, we conclude that when $\sigma_0\delta\gg1$ most of the released 
magnetic energy is converted into radiation, and the radiative efficiency can indeed 
be estimated using Eq.\ref{a13}.

\section{Radiatively cooling flow with filled gaps}
\label{sec:FCF-FG}

\begin{figure*}
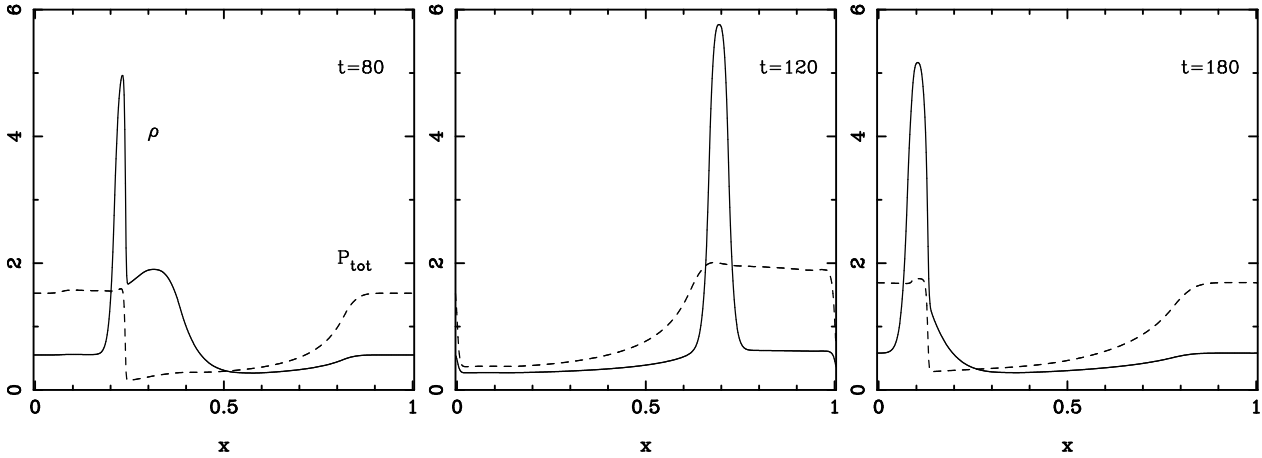

\includegraphics[width=60mm,angle=-90]{figures/mix-s80.eps}
\includegraphics[width=60mm,angle=-90]{figures/mix-s120.eps}
\includegraphics[width=60mm,angle=-90]{figures/mix-s180.eps}
\caption{Solution for the model with filled gap at $t=80,\,120,$ and 180.
The solid line shows the rest mass density and the dashed line shows the total
pressure, dominated by the magnetic pressure. As one can see, at $t=80$ and 180 
the fast forward shock crosses the flow section which has significantly higher 
rest mass density, the gap section. At the same times, the radiation loss rate, 
which is shown in right panel of Figure~\ref{fig:mix}, exhibits strong peaks.
}
\label{fig:mix-sol}
\end{figure*}

Finally, we investigate a flow with fast radiative cooling and gaps 
filled with low magnetised plasma from the start. Provided the initial state has the same 
Lorentz factor both for the gap and the pulse, the fraction of released magnetic energy 
is the same as in the chamber problem (see Eq.\ref{e10})
\begin{equation}
  -\frac{[{\cal E}_{\rm m}]}{{\cal E}_{\rm m,0}} =
  1-\frac{(1+\delta_{\rm l}^{1/2}\delta_{\rm e}^{1/2})^2}
         {(1+\delta_{\rm l})(1+\delta_{\rm e})} \, . 
\label{c1}
\end{equation}  
This energy is converted partly into the radiation and partly into the bulk kinetic 
energy. Although this partition is difficult to estimate analytically, there are no 
obvious factors that could significantly shift the balance in favour of the kinetic 
energy, suggesting that most of the magnetic energy should still be converted into 
the radiation. This is confirmed by numerical simulations

Here we present the results for the model with 
parameters $\gamma_0=5$, $\sigma_{\rm p}=10$, $\sigma_{\rm g}=0.1$, 
$\delta_{\rm l}=\delta_{\rm m}=1$, and $\delta_{\rm e}=0.01$.  
For this parameters Eq.\ref{c1} gives $[{\cal E}_{\rm m}]=-0.4{\cal E}_{\rm m,0}$.     
The left panel of Fig.\ref{fig:mix} compares the energy balance of this model with 
the empty gap model which has the same pulse parameters and hence 
$[{\cal E}_{\rm m}]=-0.5{\cal E}_{\rm m,0}$.  As one can see, in both these models 
the released magnetic 
energy does evolve towards the predicted asymptotic values and at approximately the same 
rate. The partition of this energy between the bulk kinetic energy and the radiation is 
also similar. This allows us to conclude that in the case of energetically 
subdominant gaps, the radiative efficiency is similar to what we have for empty gaps, 
and is well described by 
\begin{equation}
\eta_{\rm r} \simeq
 1-\frac{(1+\delta_{\rm l}^{1/2}\delta_{\rm e}^{1/2})^2}
        {(1+\delta_{\rm l})(1+\delta_{\rm e})} \, .
\end{equation}

Because the total accelerated mass is twice as higher in the model with filled gaps, 
one would expect the asymptotic Lorentz factor in this model to be lower compared the 
corresponding model with empty gaps. 
The data presented in Fig.\ref{fig:mix-lor} and Fig.\ref{fig:ad-rad-sol} show 
that this is indeed the case. However, the difference is rather small,  
and this suggests that the asymptotic Lorentz factor is still well described by 
Eq.\ref{b6} when the gap mass does not exceed that of the pulse.

The energy curves shown in the left panel of Fig.\ref{fig:mix} are more ragged
for the model with filled gaps. Moreover, their structure suggests a quasi-periodic 
process. In fact, the curve of radiated energy is similar in shape to those found 
in the chamber problem, indicating a strong variation of 
the dissipation rate. This is confirmed by the left panel of Fig.\ref{fig:mix}, which 
shows the energy loss rate. Just like in the chamber problem, one would expect its spikes 
to be associated with crossings of weakly magnetised gaps by shocks. 
This is indeed the case, as illustrated in Fig.\ref{fig:mix-sol}. 
This plot also shows that, like in other models, the flow develops the characteristic 
saw-tooth profile.  The time separation between these spikes
can be explained using Eq.\ref{beta-fw}. According to it, the pulse crossing time by 
a relativistic forward wave is  $\Delta t_{\rm c}\simeq 2\gamma^2 l \sim 50$, where we
used $\gamma=\gamma_0=5$. This is indeed very close to the spike separation 
in Fig.\ref{fig:mix}.      
The rapid variability inside the first spike is connected to the reverse shock, which 
crosses the shell much faster (see Eq.\ref{beta-rw}).

\section{Discussion}
\label{sec:discussion}

\subsection{Dynamics of impulsive magnetically-dominated outflows}

In recent years a great deal of progress has been achieved in the dynamics of 
magnetic steady-state flows but the observations seem to 
suggest that the central engines of cosmic jets are highly variable. 
Is this variability only a minor complication or it can
actually lead to a qualitatively different jet dynamics? The recent  
pioneering studies of this issue have suggested the latter 
\citep{C95,GKS10,L10,LL10,L11}.  However, they were mainly concerned 
with the dynamics of individual shells, presumably ejected by the central
engine, and their interaction with the external medium. These issues were
studying using rigorous mathematical modelling. On the contrary, the problem of 
interaction between  magnetically accelerated shells has remained  until now a subject of only 
rather speculative semi-quantitative  analysis.    
One of the main objectives of our study was to explore this 
important issue a bit further.  
  
One of the most important properties of the isolated shell solution is the 
concentration of energy and momentum in the compact head of the shell formed by the 
reverse rarefaction wave early on \citep{GKS10}. 
This rarefaction wave can 
be seen in the plots of the left column of Figure~\ref{fig:adiab-sol}, where it 
occupies the domain $0.45<x<0.75$. The head formation is completed when this 
rarefaction reaches the back of the shell and gets reflected as a forward 
rarefaction. The part of the shell located between the leading interface with 
vacuum and the leading front of the reflected rarefaction wave is what constitutes 
the shell head. In the source frame, it shows very little spreading in the direction 
of motion and its mean Lorentz factor increases as $r^{1/3}$. 

\citet{GKS10} and \citet{G11b} argued that this somewhat unusual, but well understood in the 
relativistic framework, behaviour of the shell head allows us to ignore the 
interaction between individual shells until their Poynting flux is almost 
fully converted into the bulk motion kinetic energy. 
Only after this, during the coasting phase, the shell spreading becomes important 
and leads to strong collisions between shells, thus allowing internal shocks 
with high dissipation efficiency. 
{\bf This assertion had apparently given a great importance to the assumption, 
made in that study, that the gaps between shells were empty, or at least so highly 
rarefied that their plasma could not influence the shell dynamics. In fact, the authors 
overlooked the ejection of shell plasma into its tail, which would rapidly fills the gap, 
and did not explore the possible implications of the head-tail collision 
for the shell dynamics. }

An isolated shell enters the coasting phase at 
$t_c\simeq (l_{\rm p}/c) \gamma_{\rm c}^2$, where $\gamma_c$ is the asymptotic Lorentz 
factor of the shell \citep{GKS10}. This is the time required for a fast magnetosonic 
wave moving relative to the shell head with the observed speed 
$ \delta v_{\rm h} = c/\gamma_{\rm c}^2$ (see Eq.\ref{beta-fw})
to traverse the head. The relative observed speed at which 
the shell plasma is ejected into the tail can be estimated via Eq.\ref{beta-rw}, which  
gives us 
$ \delta v_{\rm t} \sim 2  \sigma (\gamma_{\rm c}/\gamma)^2 \delta v_{\rm h} 
\gg \delta v_{\rm h}$.   
Thus, one can avoid the head-tail collision only if the gaps between shells 
are much wider than the shells.  

{\bf Our simulations have shown that, at least in the case there shells and gaps 
have comparable widths,  the tail-head collision strongly modifies the  
flow dynamics, making the results for an isolated shell irrelevant. The gaps soon 
become filled up with significant amounts of plasma, and thereafter we have what is best 
described as a continuous inhomogeneous flow superimposed with a wave train.  
A similar outcome is observed when the gaps are filled with
plasma already from the start.  The subsequent acceleration and radiation of this flow 
is determined by the dissipation rate of shocks, which form an integral part of 
the train,  the plasma cooling rates, 
and the flow geometry. In any case, the fraction of magnetic energy that 
can be converted into either the energy of radiation or the kinetic 
energy of the flow is set by the energy of the train. Under the condition of 
magnetic flux freezing, it is simply given by the degree of expansion in the 
longitudinal direction which the magnetised shells can achieve during the transition 
to force equilibrium and thus by the amplitude of initial fluctuations of magnetic 
pressure.}

When the radiative cooling can be ignored, the asymptotic flow parameters  
follow from the basic conservation laws of mass, magnetic flux, 
energy, and momentum. This way we find that in the case of slab geometry, the asymptotic 
Lorentz factor is $\gamma\simeq\gamma_0 (\sigma_0 \delta)^{1/2}$, rather than 
$\gamma\simeq\gamma_0 \sigma_0$ expected in the case of the total conversion of 
the magnetic energy into the bulk motion kinetic energy 
(Here $\delta=l_{\rm g}/(l_{\rm g}+l_{\rm p})$ is the relative thickness of the gap .). 
In contrast to the case of an isolated pulse \citet{GKS10}, the magnetic  
acceleration is inefficient and most of the free magnetic energy is converted into 
the thermal energy via the shock dissipation. 
However in the case of spherical geometry, the sideways expansion of the flow is 
accompanied by efficient conversion of this thermal energy into the bulk kinetic 
energy, via the thermal acceleration mechanism. This leads to the asymptotic Lorentz 
factor  $\gamma\simeq\gamma_0 \sigma_0 \delta$.  

When the radiative cooling is much faster than the adiabatic one, the difference 
between flows with slab and spherical geometry is unlikely to be significant, although
the presence of poorly radiating components such as protons complicates this issue 
and requires further investigation. Even without such components,   
the fast cooling regime is more involved as the conservation laws alone do not 
determine the asymptotic flow parameters. The outcome now depends on 
details of the radiative cooling process. The radiation reaction force may 
both help the flow acceleration and make it more difficult, depending on whether 
the photons are emitted predominantly in the forward direction in the centre-of-momentum 
frame or in the backward direction. From the numerical simulations we find that 
the emission comes from the fastest part of the flow and thus the radiation 
reaction force is a decelerating one. Most of the utilised magnetic energy 
is converted into the radiation and the asymptotic Lorentz factor of the flow 
remains significantly below the value of $\gamma_0 \sigma_0 \delta$, characteristic 
of efficient magnetic acceleration.

\subsection{Astrophysical implications}

In terms of astrophysical implications, our results are most relevant to the issue 
of internal shock dissipation and emission from relativistic jets of GRBs and AGN.   
A detailed test of the shock dissipation model in magnetically-dominated flows 
against observations is beyond the scope of this paper. Here we only outline 
few issues to be investigated in future studies.     

Just like in the hydrodynamic model, the energy reservoir for the internal shock
dissipation in the magnetic model is associated with the variable component of 
the jet. For Poynting-dominated jets this is the magnetic energy of 
the fast magnetosonic waves  driven into the jet by the variable central engine.
Their contribution to the overall jet energy budget depends on 
the exact details of the central engine operation, and at least in principal it may 
well be dominant. 

Provided the radiative cooling time is sufficiently short, most of the 
dissipated energy is converted into radiation. In our simplified test problems, 
the radiative efficiency is given by the Equation~\ref{e10}, which shows that it 
can be very substantial even for a rather moderate central engine variability. 
Thus as far as the observed radiative efficiency of GRB jets is concerned, 
the internal shock model of prompt GRB emission and the magnetically-dominated 
model of GRB jets are not mutually excluding.  
However, the dissipation length scale increases with the jet magnetization and may 
exceed the jet length.

In the hydrodynamical version of the internal shock model, the prompt emission 
originates from the region where individual ballistic shells collide with each 
other because of the differences in the ejection speed \citep{Piran}. For strong 
variation of the Lorentz factor, this occurs at 
\begin{equation}
R_{\rm s}\simeq\gamma_{\rm j}^2 c \delta t_{\rm s} 
\simeq 3\times10^{16} \mbox{cm} \fracb{\gamma_{\rm j}}{10^3}^2 
\fracb{\delta t_{s}}{\mbox{s}} \, ,
\label{R_s}
\end{equation}
where $\delta t_{\rm s}$ is the 
characteristic time interval between the shell ejections\footnote{Such a characteristic time 
does not have to exist. Instead, the central engine may exhibit a wide distribution for 
the shell ejection time.}, and $\gamma_{\rm j}$ is the Lorentz factor of the slower 
shell. Each collision gives rise to a pulse of the prompt emission light-curve. 
Its duration, $\delta t_{\rm p}$, is determined by the curvature of the 
shock front and the Doppler beaming (the so-called ``angular spreading effect''), 
and it turns out to be the same as $\delta t_{\rm s}$ \citep{Piran}.

In the magnetic model, the prompt emitting region can extend well beyond $R_{\rm s}$. 
According to Eq.\ref{t_relax} a forward shock dissipates its energy when it covers 
the distance  $R\simeq\sigma R_{\rm s}$. For $t_{\rm p}\sim 1\,$s, $R_{\rm s}$ 
is already dangerously close to the radius of the external shock which seems 
to rule out $\sigma\gg 1$. However, the problem arises only if we associate 
each strong individual pulse of a GRB light curve with an individual shock wave. 
If instead each such pulse is associated with a whole packet of shocks, 
so that $\delta t_{\rm s} < \delta t_{\rm p}/\sigma$, this is no longer an issue. 
Obviously, this explanation implies a secondary central engine variability process,
operating on the time $\delta t_{\rm p}$, which modulates the output of the 
primary process, operating on the timescale $\delta t_{\rm s}$.  
In fact, this way one can explain why about $20\%$ of GRBs have rather featureless 
light curves \citep{Piran}. This may just be the case of weak modulation.  

In this model, there are $\sim\sigma$ shocks inside the dissipation zone and one 
may wonder if their emission signals overlap. In the source frame 
the relative speed of a light signal and a forward fast magnetosonic wave propagating 
in the radial direction is 
\begin{equation}
\delta v_{\rm \gamma f} \simeq \frac{c}{8\sigma\gamma_{\rm j}^2} \, .
\end{equation}
The travel time across the dissipation zone is $\sigma R_{\rm s}/c$. During this 
time the distance between these two waves is changed by only 
$c\delta t_{\rm s}/8$. Thus, the overlapping could only be a result of the angular 
spreading effect, which spreads $\delta$-shape signals over the time  
$\delta t_{\rm ang} = R/2c\gamma_j^2$, where $R$ is the emission radius.  
This gives us  $\delta t_{\rm ang} = \delta t_{\rm s}/2$  at the beginning of the 
dissipation zone ($R\sim R_s$) and 
$\delta t_{\rm ang} = \sigma \delta t_{\rm s}/2 \gg \delta t_{\rm s}$ 
at the end of the dissipation zone ($R\sim \sigma R_s$). One may draw two conclusions 
from these numbers. First, a micro-pulse from an individual shock must have a long 
smooth tail of length $\simeq \sigma \delta t_{\rm s}/2$. 
Spike-like features on the scale of $\sim \delta t_{\rm s}$ can be found only at the head 
of this micro-pulse  (They may correspond to shock crossings of contact 
discontinuities in the jet; see Fig.\ref{fig:dedt} and Fig.\ref{fig:rad-energy}).   
Second, the tails of individual micro-pulses overlap but their leading spikes
do not and can give rise to a pulse substructure. The fact that, shocks are 
strongest at the beginning of the dissipation zone increases chances of them being
detected.  This could be the origin of the observed variability of GRBs on 
millisecond time scale \citep{WS00}.

How short can $\delta t_{\rm s}$ be? 
In the Blandford-Znajek model of GRB central engine, the only ``easy'' way of 
changing its jet power, $L_{\rm j}$, is via changing the flux of open magnetic 
field lines,  $\Psi$, which threads the black hole, as these parameters relate via  
$ L_{\rm j} \propto \Psi^2$ \citep{BZ77}.  This magnetic flux may change 
significantly if the disk drags in an alternating magnetic field, which may even
bring about a change of polarity of the black hole magnetosphere\footnote{  
The effective loss of magnetosphere and its shielding action during 
a change of polarity may also let weakly magnetized surrounding plasma 
to enter the jet channel and to become entrained by the jet.}. 
The relevant time scale for this process is probably 
the disk inner edge  ``viscous'' time scale. For an $\alpha$-disk, this is    
\begin{equation}
\delta t_{\rm v} \approx 10
\left(\frac{\alpha\,\delta^2}{10^{-3}}\right)^{-1}
\left(\frac{M}{M_{\odot}}\right)\;\mbox{ms}\ ,
\label{tmin}
\end{equation}
where $M$ is the black hole mass and $\delta = H_d/R_d$ is the ratio of the disk 
height to its radius \citep{SHS73}.  This is almost a hundred times shorter 
compared to the observed typical separation between GRB pulses \citep{N96}.  
horizon. The shortest scale for restructuring of the black hole magnetosphere 
is given by the light crossing time of the ergosphere 
\begin{equation}
\delta t_{\rm er} =  \frac{2GM}{c^3}  \approx 10 
\left(\frac{M}{M_{\odot}}\right)\;\mu\mbox{s}\ , . 
\label{tmin1}
\end{equation}
A rapid and frequent restructuring of magnetosphere may also be typical for a 
newly-born millisecond magnetar, due to its ultra-strong magnetic field, lack of 
solid crust, and active magnetic dynamo. These estimates assume fast magnetic 
reconnection in the magnetospheres, which has been questioned recently on the 
basis of collisional nature of reconnection of super-strong magnetic field 
\citep{MU10,LM11,U11}, although even the collisional magnetic reconnection can be 
fast due to the secondary tearing instability 
\citep[see the discussion and references in ][]{U11}.  

As to the modulation process, it should also involve variation of $\Psi$ but on 
a larger time scale. For example, more massive accretion disks could support stronger 
magnetic field and result in larger magnetic flux trapped by the black hole. 
Variations of the disk mass could result from unsteady stellar collapse.     
The typical duration of strong GRB pulses is around the free-fall time  
from the radius of $\sim 10^9$cm for the GRB progenitor in the collapsar model. 
Given the total radius of the progenitor, $\sim R_\odot$, it does not seem implausible 
for the GRB jet to perturb the stellar mass distribution on this scale.  

{\it Swift} observations of early X-ray afterglows have discovered the
presence of plateaus as well as strong flares in their light-curves,
both unexpected in the original external shock model \citep{Z07b}.
One possible and perhaps the most likely explanation of these features is
that they represent the ``late prompt emission'' of a long-lasting central engine
\citep{GGNF07,GNGC09}. This interpretation is supported by a number of similarities
between the X-ray flares and the GRB sub-pulses \citep{M10}.
Such a long-lasting activity, at least up to $10^4$ seconds,
is inconsistent with the high mass accretion rate,
above $10^{-2}M_\odot/$s, required by the neutrino annihilation mechanism of
jet production and implies the magnetic mechanism.

In addition to the similarities between the X-ray flares and the
GRB sub-pulses, there are also few differences \citep{M10}.
For example, the flare duration increases linearly with the flare time since GRB
trigger.  In our model, the duration of prompt pulses is determined by the duration 
of strong accretion episodes in the history of the central engine. If the X-ray 
flares are associated with the fallback accretion, then the flare time is 
likely to be determined by the location of fallback turning point whereas its 
duration by the spatial dispersion of falling back material. They may well 
correlate with each other.   

The evolution of individual pulses of GRBs is often described as 
``Fast Rise Exponential Decay'' \citep[FRED, e.g.][]{Piran}. The latest results show that
on average both the prompt pulses and the X-ray flares have approximately 
twice as shorter the rise time compared to the decay time \citep{N05,C10}. 
Although shorter rise times are expected in the standard internal shock model, 
it does not really say by how much \citep{Yi04}. Perhaps, this tells us that 
the pulse shape is not determined by the shell collision after all. Other 
processes may have similar time properties. In fact, FRED mass accretion rate 
could be a general consequence of the diffusive transport in episodic accretion 
disks \citep{W01}. 
 
We have shown that, provided the GRB jets are cooled mainly radiatively, 
they are likely to remain Poynting-dominated even beyond the prompt emission 
zone. The exact nature of piston driving the 
external shock into the surroundings of GRB has little effect on the shock 
dynamics and emission. Provided the total energetics is the same,  
both the kinetic energy dominated and the magnetic energy dominated 
ejecta produce the same afterglow emission associated with this  
shock \citep{LB03,L05,L11,MGA09}. However, the reverse shock, which is driven into 
the ejecta, will be much less dissipative if the ejecta is highly magnetized 
\citep{ZK05,MGA09}. This could be the reason behind the paucity of the 
lower energy, most likely optical, flashes expected from the reverse shock 
in the non-magnetic models of afterglows \citep{G09}.

As to the issue of the jet acceleration, our results  
indicate that the impulsive magnetic acceleration mechanism, proposed by \citet{GKS10}, 
is unlikely to operate in its original form. Even if we ignore the entrainment of 
surrounding plasma, the strong shock interaction between heads and tails of magnetic 
shells, revealed in the present study, is already sufficient to prevent the conversion 
of almost all available magnetic energy into the kinetic energy shell heads. Instead, 
the problem reduces to its more standard form where non-linear waves travel along the 
jet, interacting with the mean flow. The efficiency of the mean flow acceleration by 
these waves depends on the efficiency of its radiative cooling, and hence on the photon 
opacity of the flow. If this cooling is weak then the jet acceleration can be very 
efficient, with a large fraction of the wave Poynting flux converted into the kinetic 
energy of the flow. Otherwise, it is converted into radiation. Based only on baryonic 
electrons the optical thickness of a GRB jet to Thomson scattering is 
\begin{equation} 
    \tau \simeq 1 \frac{L_{52}}{r_{13}\sigma\gamma_{\rm j,2}^3}\, ,
\end{equation}
where $L$ is the isotropic jet power. Thus, it seems unlikely for the dissipation 
zone to be optically thick, unless pairs dominate the opacity. This agrees with  
the high observed radiative efficiency of GRB jets.   

Impulsive jet production provides favourable conditions for entrainment of weakly 
magnetised plasma that may exist in the vicinity of the central engine.  We focused 
on the implications of this process for the dissipation efficiency of shock waves 
in magnetically dominated flows. But in addition to this, the presence of weakly 
magnetised plasma may also help to overcome another difficulty of the shock model, 
related to the shock acceleration of non-thermal particles. 
However, for this to work the magnetization may have  
to be very low, down to $\sigma<10^{-3}$ \citep{SS10}.

\section{Conclusions}
\label{sec:conclusions}

In this paper we analysed the potential role of shock dissipation on the 
dynamics and emission of impulsive Poynting-dominated relativistic jets.
The main insights came from analytical and numerical solutions of Relativistic MHD 
equations in slab geometry. For computational reasons the numerical simulations were 
limited to flows with Lorentz factors which were much lower compared to those deduced 
from the observations of GRBs. 
These and other limitations of this study warn against making firm conclusions and 
a certain degree of uncertainty definitely remains in many respects, and particularly 
when it comes to astrophysical applications. Keeping this in mind,  
our main conclusions are  
   
\begin{itemize}
\item 
The dissipation efficiency of strong shocks in highly magnetized plasma 
is low $\simeq 0.5 (1+\sigma)^{-1}$, mainly because only the kinetic energy 
dissipates and it represents only a small fraction of the total energy flowing 
through the shock. 
For moderate magnetization, $\sigma\simeq 1$, the shock dissipation efficiency
is reasonably high, $\simeq 30$\% of the total energy flux. 
\item
The dynamics of a magnetic shell in a train of shells, all ejected in the same 
direction and initially separated by empty space, can be rather different from 
that of an isolated shell in vacuum, studied in \citet{GKS10}. 
Unless the separation between these shells is very 
large, they strongly interact with each other. The result of this interaction 
is best described as an underlying continuous flow superimposed with strong
magnetosonic waves travelling in the same direction. The wave Poynting flux 
provides energy for heating and acceleration of the flow. 
A similar outcome is observed in models where gaps are filled with 
energetically sub-dominant plasma from the very beginning. 
\item
For an infinite flow, the reservoir of magnetic energy is given simply by the 
decrease of the magnetic energy, as dictated by the magnetic flux freezing, 
during the transition to a wave-free final state. 
The shock dissipation is an essential part of this transition.  
The radiative cooling allows to maximise the released magnetic energy. 
For radially diverging flows, the adiabatic cooling is expected to do the same.      
In both cases, the fraction of released magnetic energy does not depend on 
the initial flow magnetization but only on the initial magnetic inhomogeneity of 
the flow,  
or in other words on the fraction of energy carried by the waves.   
However, the flow magnetization determines the tempo and the characteristic 
length scale of the energy release, which grows $\propto\sigma$ for $\sigma\gg 1$. 
\item
With application to GRB jets, this shows that the internal shocks may still 
be responsible for the prompt emission even in the case of Poynting-dominated 
jets. At least,
the high observed fraction of the prompt emission in the total energy 
budget of GRB events is not inconsistent with this model. 
\item
However, the increased size of the prompt emission (internal shock dissipation) zone, 
compared to that of the non-magnetic model, calls for a somewhat different interpretation 
of the prompt emission variability. In particular, the observed individual pulses with duration 
around one second are unlikely to be associated with the emission of individual shocks,
as this would push the dissipation zone  beyond the external shock radius.
Instead, each such relatively long pulse could represent the combined emission 
from a whole pack of shocks associated with mini-shells ejected by the central 
engine on a much shorter timescale.  The long 
timescale variability would then be related to some modulation process, determining 
the energy of emitted mini-shells.  In fact, both the magnetar and black hole 
magnetospheres do allow variability on millisecond,  and possibly even shorter, 
timescales. As to the nature of the modulation, one possible mechanism is the unsteady 
mass supply to the accretion disk in the collapsar scenario.      
\item
The emission from individual shocks may produce fine substructure, 
on the timescale of shell ejection. This could be the origin of the observed 
millisecond spikes in the light-curves of the prompt emission.     
\item
After leaving the zone of internal dissipation, the initially Poynting-dominated jet
is most likely to remain Poynting-dominated. This may not effect the dynamics 
of the forward shock and its afterglow emission, but will result in reduced emission 
from the reverse shock compared to the non-magnetic model. Yet, the presence of 
weakly magnetized domains in  the jet 
will increase the radiative efficiency of this shock compared to the value expected  
for a uniform jet with the same mean magnetization.    

\end{itemize}

\section*{Acknowledgments}

The author is grateful to J.McKinney, Y.Lyubarsky, 
M.Lyutikov, the anonymous referee, and particularly to J.Granot for 
helpful discussions and constructive criticism of the original manuscript.

\appendix

\section{Relativistic Perpendicular MHD Shocks}
\label{appendix-1}

Here we analyse MHD shock waves in the special case where the 
magnetic field is parallel to the shock front and the flow velocity 
is perpendicular to it.  
In the shock frame, the fluxes of energy, momentum, rest mass, and 
magnetic field are continuous across the shock 
\begin{equation}
\label{s1}
(w+B^2)\gamma^2 v = \mbox{const},
\end{equation}
\begin{equation}
(w+B^2)\gamma^2 v^2 +p +\frac{B^2}{2} = \mbox{const},
\label{s2}
\end{equation}
\begin{equation}
\rho\gamma v = \mbox{const},
\label{s4}
\end{equation}
\begin{equation}
B\gamma v = \mbox{const},
\label{s5}
\end{equation}
where $\rho$ is the rest mass density, $p$ is the gas pressure, $w=\rho + \kappa p$ is 
the relativistic enthalpy, $\kappa=\Gamma/(\Gamma-1)$, where $\Gamma$ is the adiabatic 
index, $B$ is the magnetic field as measured in the fluid frame,   
and $\gamma$ is the Lorentz factor. 
We select the frame where the velocity vector is normal to the shock plane and 
the magnetic field is parallel to it.  We use subscripts 1 and 2 to denote the 
upstream an the downstream states respectively. 

Equations (\ref{s1}),(\ref{s4}) and (\ref{s5}) yield 
\begin{eqnarray}
\frac{\rho_2}{\rho_1}=\frac{B_2}{B_1}=\frac{\sigma_2}{\sigma_1}=\frac{\delta}{\chi},
\label{os2}
\end{eqnarray}
\begin{equation}
a_2^2=\frac{1}{\kappa} \left(  
  \delta(1+\kappa a_1^2 + \sigma_1)-\sigma_1\fracb{\delta}{\chi} -1
\right) \, , 
\label{os3}
\end{equation}
where 
\begin{equation}
\sigma = B^2/\rho
\end{equation}
is the magnetization parameter, 
\begin{equation}
a^2=p/\rho
\end{equation}
is the temperature parameter, 
\begin{equation}
\chi=v_{2}/v_{1},
\end{equation}
\begin{equation}
   \delta=\frac{\gamma_1}{\gamma_2}=(1+u_{1}^2(1-\chi^2))^{1/2},
\label{s6}
\end{equation}
and $u=v\gamma$. 
Using the definitions of $a$ and $\sigma$, Equation (\ref{s2}) can be written as 

\begin{eqnarray}
\nonumber
(1+\kappa a_1^2+\sigma_1)\frac{j^2}{\rho_1} + \rho_1(a_1^2+\frac{1}{2}\sigma_1) = 
\end{eqnarray}
\begin{equation} 
\qquad\qquad
(1+\kappa a_2^2+\sigma_2)\frac{j^2}{\rho_2} + \rho_2(a_2^2+\frac{1}{2}\sigma_2), 
\label{os4}
\end{equation}
where $j=\rho\gamma\beta$. Via substituting the expressions for $\rho_2$, 
$\sigma_2$, and $a_2$ from  Eqs.(\ref{os2},\ref{os3}) this equation becomes 
an equation for $\chi$, which defines it as a function of the upstream state 
parameters.  
In general, this is a rather combersome algebraic equation which has to be 
solved numerically. However, in many astrophysical applications one may 
assume that the upstream state is cold ($a_1 \to 0$), which allows significant 
simplifications. Then Eq.\ref{os4} reduces to  
\begin{equation}
  a_3 \chi^3 + a_2 \chi^2 + a_1 \chi +a_0 + g(\chi) = 0\, , 
\label{s7}
\end{equation}
where 
\begin{eqnarray}
\nonumber
a_3 = (1+\sigma_1)u_{1}^2\fracb{\kappa-1}{\kappa}\, , 
\end{eqnarray}
\begin{eqnarray}
\nonumber
a_2 = -\sigma_1u_{1}^2 \fracb{\kappa-2}{2\kappa} \, , 
 -(1+\sigma_1)u_{1}^2 -\frac{\sigma_1}{2} 
\end{eqnarray}
\begin{eqnarray}
\nonumber
a_1 = \frac{(1+\sigma_1)(1+u_{1}^2)}{\kappa} \, , 
\end{eqnarray}
\begin{eqnarray}
\nonumber
a_0 = \sigma_1 \fracb{\kappa-2}{2\kappa} (1+u_{1}^2) \, , 
\end{eqnarray}
and $g(\chi)= -\delta(\chi)\chi/\kappa$. One can see that 
the solutions of this equation are parametrised only by the upstream 
magnetization and the shock speed.  
Instead of the shock speed one can introduce the more traditional 
shock Mach number, that will be done later.   

For any values of upstream parameters Eq.\ref{s7} must allow the 
trivial continuous solution, $\chi=1$, which describes a flow without a shock. 
Thus, we have  
\begin{equation}
  a_3 (\chi^3-1) + a_2 (\chi^2-1) + a_1 (\chi-1) =g(1)-g(\chi), 
\end{equation}
which can also be written as 
\begin{equation}
  a_3 \chi^2 + (a_3+a_2) \chi + (a_3+a_2+a_1) =\frac{f(\chi)}{\kappa}, 
\label{s8}
\end{equation}
where 
$$
    f(\chi) = \frac{g(1)-g(\chi)}{\chi-1} = 
    \frac{1-\chi\delta(\chi))}{(1-\chi)}.
$$ 
For this trivial solution both the numerator and the denominator in 
the above expression for $f(\chi)$ vanish. 
An additional continuous solution, $\chi=1$, may  
exist when the shock speed relative to the upstream 
state equals to the fast magnetosonic speed of this state. 
In order to verify this, one can replace $f(\chi)$ in Eq.\ref{s8} 
with its limiting value 
$$
  \lim_{\chi\to 0} f(\chi) = 1-u_1^2
$$    
and then substitute $\chi=1$ into the left-hand side of this equation. 
The result is 
\begin{equation}
   u_1^2=\sigma_1,
\label{s9}
\end{equation}
which is the fast magnetosonic condition. Indeed, in the limit of 
cold flow, the fast magnetosonic speed is isotropic and equals to the 
Alfv\'en speed 
\begin{equation}
c_{\rm f}^2 = \frac{B^2}{B^2+\rho} = \frac{\sigma}{1+\sigma}.
\label{s10}
\end{equation}
The corresponding Lorentz factor $\gamma_{\rm f}=\sqrt{1+\sigma}$.
Thus the condition (\ref{s9}) becomes the condition 
\begin{equation}
M_{\rm 1} = 1
\end{equation}
on the shock fast magnetosonic Mach number, which is defined as  
\begin{equation}
M_{\rm 1} = \frac{u_1}{u_{\rm f,1}}= \frac{u_1}{\sqrt{\sigma_1}},
\label{s11}
\end{equation}
where $u_{\rm f}=c_{\rm f}\gamma_{\rm f}$. With $M_{\rm 1}$ and
$\sigma_1$ as parameters, Eq.\ref{s8} reads
\begin{equation}
  a \chi^2 + b \chi + c =2 f(\chi), 
\label{s12}
\end{equation}
where 
\begin{eqnarray}
\nonumber
a = 2(1+\sigma_1)\sigma_1(\kappa-1) M_{\rm 1}^2, 
\end{eqnarray}   
\begin{eqnarray}
\nonumber
b = -\sigma_1(2+\kappa\sigma_1) M_{\rm 1}^2-\kappa\sigma_1, 
\end{eqnarray}   
\begin{eqnarray}
\nonumber
c = -\sigma_1^2(\kappa-2) M_{\rm 1}^2-\kappa\sigma_1-2(\sigma_1+1). 
\end{eqnarray}   

For  $M_{\rm 1}\gg \max(1,1/\sigma_1,1/\sqrt{\sigma_1})$ 
one can only retain the terms that are proportional to  $M_{\rm 1}^2$ 
in the coefficients of $\chi^i$. This yields 
\begin{equation}
  2(1+\sigma_1)(\kappa-1) \chi^2 - (2+\kappa\sigma_1) \chi + 
  (2-\kappa)\sigma_1 = \frac{2}{M_{\rm 1}^2} f(\chi). 
\label{s13}
\end{equation}
The factor $M_{\rm 1}^{-2}$ on the right-hand-side term is small, 
suggesting that this term can be also ignored. This gives us 
simple quadratic equation which does not involve $M_{\rm 1}$,  
\begin{equation}
  2(1+\sigma_1)(\kappa-1) \chi^2 - (2+\kappa\sigma_1) \chi + 
  (2-\kappa)\sigma_1 =0. 
\end{equation}
Since, the physical meaning of $\chi$ dictates $0<\chi\le 1$, 
the only suitable solution of this equation is 
\begin{equation}
  \chi = \frac{2+\kappa\sigma_1 +\sqrt{D}}{4(\kappa-1)(1+\sigma_1)},
\label{s14}
\end{equation}
where
$$
  D=(2+\kappa\sigma_1)^2-8(1+\sigma_1)\sigma_1(\kappa-1)(2-\kappa).
$$
This is the same solution as in the analysis of oblique shocks in 
\citet{KL11} and, for $\kappa=4$ ($\Gamma=4/3$), in the perpendicular 
shock solution of \citet{KC84}.

Further simplification is possible when $\sigma_1 \gg 1$. 
In this case 
\begin{equation}
  \chi \simeq 1 + \frac{4-2\kappa}{3\kappa-4} \sigma_1^{-1},
\end{equation}
which for $\kappa=4$ gives us 
\begin{equation}
  \chi \simeq 1 - \frac{1}{2} \sigma_1^{-1}. 
\end{equation}
Using this result one can find  
\begin{equation}
   \gamma_2 \simeq \sigma_1^{1/2},
\end{equation}
\begin{equation}
   \rho_2 \simeq M_{\rm 1} \rho_1,
\end{equation}
\begin{equation}
   p_2 \simeq \frac{1}{8} \rho_1 M_{\rm 1}^2,
\end{equation}
\begin{equation}
   B_2 \simeq  M_{\rm 1} B_1,
\end{equation}
\begin{equation}
   p_{\rm m,2} \simeq   M_{\rm 1}^2 p_{\rm m,1},
\end{equation}
\begin{equation}
   \beta_{2} \simeq  \frac{1}{4}\sigma_1^{-1},
\end{equation}
\begin{equation}
   \sigma_{2} \simeq  M_{\rm 1} \sigma_1.
\end{equation}
These allow to find how energy is distributed between its different forms 
in the downstream flow. 
Denoting as $F_{\rm k}=\rho\gamma^2v$, $F_{\rm m}=B^2\gamma^2v$, 
and $F_{\rm t}=\kappa p\gamma^2v$ the fluxes of kinetic, magnetic, and thermal 
energy respectively, we have 
\begin{equation}
   F_{\rm k,2} \simeq \frac{1}{M_{\rm 1}}F_{\rm k,1}, 
\end{equation}   
\begin{equation}
   F_{\rm t,2} \simeq \frac{1}{2}F_{\rm k,1},
\end{equation}
and 
\begin{equation}
   F_{\rm m,2}-F_{\rm m,1} \simeq \frac{1}{2}F_{\rm k,1}.
\end{equation}
Thus, approximately one half of the upstream kinetic energy 
is dissipated into heat, whereas the other half is converted 
into the magnetic energy.   
The shock dissipation efficiency can be defined as 
\begin{equation}
   \eta_{\rm s} = \frac{F_{\rm t,2}}{F_{\rm tot}},  
\end{equation}
where $F_{\rm tot}= F_{\rm m}+F_{\rm k}+F_{\rm t}$ is the total energy flux. From the 
above results it follows that
\begin{equation}
   \eta_{\rm s} \simeq \frac{1}{2(1+\sigma_1)}. 
\end{equation}
Thus, the shock dissipation efficiency is greatly reduced in highly 
magnetised plasma. 

When $\sigma_1\gg 1$ we have $f(\chi)\simeq -2\sigma_1 M_{\rm 1} $. 
Substituting this result into Eq.\ref{s13}, one can verify that 
the right hand side term of this equation is indeed much smaller 
compared to any term on the left hand side. 

We should point out 
that the condition $u_1\gg 1$, stated in 
\citet{KC84} for their approximate shock solution, is not quite correct. 
The proper condition which leads to the solution 
(\ref{s14}) is  $M_{\rm 1}\gg 1$ and it requires $u_1\gg\sqrt{\sigma_1}$.        
The shock solution does not even exist when $M_{\rm 1} \le 1$ whereas  
the condition $u_1\gg 1$ may still be satisfied. 

\citet{ZK05} use the Lorentz factor of the relative motion between 
the states on both sides of the shock, 
$$
\gamma_{12} = \gamma_1 \gamma_2 (1-v_1v_2),
$$
as a shock strength parameter. Using the above results, it is 
easy to show that for $ M_{\rm 1}\gg 1$ and $\sigma_1\gg 1$
$$
 \gamma_{12} \simeq \frac{M_{\rm 1}}{2},
$$   
and hence this parameter is equivalent to the shock Mach number. 
However, for lower  $ M_{\rm 1}$ and $\sigma_1$ the connection between 
these parameters is more involved. The shock Mach 
number is a traditional parameter which is rightly recognised as
most useful in characterising the shock strength. It describes 
the state upstream of the shock, leaving all the downstream state 
parameters to be found from the shock equations, in contrast to 
$\gamma_{12}$, which involves both these states.


\end{document}